\documentclass[lettersize,journal]{IEEEtran}
\usepackage{amsmath,amsfonts}
\allowdisplaybreaks[1] 
\usepackage{algorithmic}
\usepackage{algorithm}
\usepackage{array}
\usepackage[export]{adjustbox} % 必须引入! 提供 valign=c 让图片垂直居中
\usepackage{textcomp}
\usepackage{color} 
\usepackage{stfloats}
\usepackage{url}
\usepackage{xcolor} % 导言区
\usepackage{verbatim}
\usepackage{graphicx}
\usepackage{cite}
\usepackage{bm}  % 导言区加上
\usepackage{subfigure}
\begin{document}

\title{Airy Beam Engineering in Near-field Communications: \\ A Tractable Closed-Form Analysis in the Terahertz Band}

\author{Wenqi Zhao, Chong Han,~\IEEEmembership{Senior Member,~IEEE},
and Emil Bj{\"o}rnson,~\IEEEmembership{Fellow,~IEEE}

        % <-this % stops a space
\thanks{
    Wenqi Zhao is with the Terahertz Wireless Communications (TWC) Laboratory, Shanghai Jiao Tong University, Shanghai 200240,
    China (e-mail: wenqi.zhao@sjtu.edu.cn).

    Chong Han is with Terahertz Wireless Communications (TWC) Laboratory and also the Cooperative Medianet Innovation Center (CMIC), School of Information Science and Electronic Engineering, Shanghai Jiao Tong University, Shanghai 200240,
    China (e-mail: chong.han@sjtu.edu.cn).

  E. Bj{\"o}rnson is with the Department of Communication Systems, KTH Royal Institute of Technology, Stockholm 10044, Sweden (email: emilbjo@kth.se).
   
}}


 % \markboth{IEEE Transactions on Wireless Communications, Submitted in March 2026}
 % {Shell \MakeLowercase{\textit{Zhao et al.}}: Closed-form Airy Beam Design.}

%column for its text to clear the IEEEpubid mark.

\maketitle

\begin{abstract}

Terahertz (THz) communication can offer terabit-per-second rates in future wireless systems, thanks to the ultra-wide bandwidths, but require large antenna arrays. As antenna apertures expand and we enter the near-field scenarios, the conventional binary classification of communication links as either Line-of-Sight (LoS) or Non-Line-of-Sight (NLoS) becomes insufficient. Instead, quasi-LoS scenarios, where the LoS path is partially obstructed, are increasingly prevalent, posing significant challenges for traditional LoS focusing and steering beams. The Airy beam serves as a promising alternative, utilizing its non-diffracting and curved trajectory properties to mitigate such blockages. However, while existing electromagnetics literature primarily explores their physical patterns without practical generation schemes, recent communication-oriented designs predominantly rely on learning-based frameworks lacking interpretable closed-form solutions.
To address this issue, this paper investigates a closed-form Airy beam design to efficiently synthesize Airy beam phase profiles based on the positions of the transceivers and obstacles. Specifically, rigorous analytical derivations of the electric field and trajectory are presented to establish a deterministic closed-form design for ULA Airy beamforming. Leveraging 3D wavefront separability, this framework is extended to uniform planar arrays (UPAs) with two operation modes: the hybrid focusing-Airy mode and the dual Airy mode. Simulation results verify the effectiveness of our derived trajectory equations and demonstrate that the proposed closed-form design significantly outperforms conventional beamforming schemes in quasi-LoS scenarios. Furthermore, the proposed method achieves performance comparable to exhaustive numerical searches with low computational complexity and enhanced physical interpretability.

\end{abstract}


\section{Introduction}
\IEEEPARstart{W}{ith} the unfolding vision of future communication systems, the requirements for wireless communication rates and latency have reached unprecedented heights~\cite{Akyildiz-2022-Terahertz,Chen-2019-A-survey}. To satisfy the demand for ultra-high data transmission rates at the terabits per second (Tbps) level, the terahertz (THz) frequency band, offering tens of GHz or even wider available bandwidth, is envisioned as a key enabling technology. However, to maintain a sufficient link budget when the individual antennas shrink, Ultra-Massive Multiple-Input Multiple-Output (UM-MIMO) technology has been widely adopted, employing hundreds or even thousands of antennas to generate high-gain directional beams~\cite{Han-2021-Hybrid}. As the antenna aperture size increases and the operating frequency rises, the Rayleigh distance of the communication system extends significantly, placing the transceivers predominantly within the near-field region rather than the conventional far-field region~\cite{An-2024-Near-Field}.
Specifically, for a fixed aperture length of $0.4$~m, the Rayleigh distance scales from $3.2$ m at $3$~GHz to $32$ m at $30$~GHz, reaching an extensive $320$~m at $300$~GHz. This shift ensures that THz systems predominantly operate within the near-field region. Consequently, this transition inherently shifts the wave model from plane wave in lower frequency bands to spherical wave in THz near-field scenarios, simultaneously unlocking new degrees of freedom for advanced wavefront shaping.

Despite the unprecedented potential of THz communications, the links are highly vulnerable to blockage by obstacles. In practical communication scenarios, due to the enlarged array aperture, the channel environment is no longer limited to ideal Line-of-Sight (LoS) or completely blocked Non-Line-of-Sight (NLoS) conditions. Instead, partially blocked LoS scenarios~\cite{Yuan-2023-Spatial}, referred to as quasi-LoS, have emerged especially in complex indoor scenarios. A prominent example is the wireless data center~\cite{Han-2025-datacenter-mag}, a key application scenario for THz communications, where dense server racks, structural walls, and metallic mesh antenna arrays substantially raise the probability of partial LoS blockage, with up to 52\% partial obstruction. Consequently, such quasi-LoS scenarios, lying between LoS and NLoS, are becoming increasingly prevalent.
%In these scenarios, the presence of obstacles truncates the energy of traditional beams, causing a sharp decline in signal strength and severely impairing communication quality.

Facing this challenge, existing LoS beam patterns suffer from severe limitations under quasi-LoS conditions. As shown in Table~\ref{tab:beam_comparison}, traditional far-field steering beams and near-field focusing beams, which are fundamentally Gaussian beams characterized by an amplitude profile following a Gaussian distribution~\cite{Ayach-2014-Spatially,Alkhateeb-2015-Limited,Zhang-2022-Beam-focusing,Cui-2022-Channel-estimation,Cui-2023-Near-Field-MIMO,Wu-2023-Multiple}, perform excellently in unobstructed LoS paths but suffer significant performance degradation in quasi-LoS scenarios due to their inherent inability to circumvent blockages. On the other hand, although utilizing environmental scattering or reflection to construct NLoS paths serves as an alternative, the reflection loss in the THz band is extremely high, often failing to provide the sufficient signal power to support high data rates. To address this issue, recent studies have proposed utilizing Intelligent Reflecting Surfaces (IRSs) to construct additional NLoS paths to circumvent obstacles~\cite{Chen-2021-Towards,Dutin-2025-Non-Line-of-Sight}. Although an IRS can effectively mitigate blockage problems, its deployment typically incurs additional hardware costs, complex channel state information acquisition, and phase synchronization overheads. Furthermore, in space-constrained or dynamically changing environments, the conditions for deploying IRS are not always available.

Therefore, existing Gaussian beamforming schemes are ineffective in such scenarios, making the exploration of alternative beamforming strategies for near-field quasi-LoS environments imperative. As illustrated in Table~\ref{tab:beam_comparison}, distinct from the Gaussian beam with a quadratic polynomial phase profile, the Airy beam is characterized by a cubic phase profile and a transverse field following the Airy function. It exhibits unique self-acceleration, and non-diffraction properties, which have garnered widespread attention. Specifically, its non-diffraction characteristic allows the beam to maintain high energy concentration over long distances without significant spatial dispersion. Simultaneously, the self-healing property enables the wavefront to reconstruct itself even after the main lobe encounters partial blockage , ensuring a robust and reliable link in complex THz near-field environments. These characteristics position the Airy beam as a highly promising solution for mitigating the detrimental effects of quasi-LoS obstruction in THz communications.

\begin{table*}[htbp]
    \centering
    \caption{Comparison of Beamforming Schemes under Different Communication Scenarios}
    \label{tab:beam_comparison}
    
    % 1. 设置行高，让文字部分不拥挤
    \renewcommand{\arraystretch}{1.3} 
    % 2. 设置列宽，让表格稍微宽敞一点
    \setlength{\tabcolsep}{6pt}

    % 注意：使用了 | 竖线，就不要用 booktabs 的 \toprule 等命令，否则线会断
    \begin{tabular}{|l|c|c|c|}
        \hline 
        \textbf{Scenario} & 
        \text{\shortstack{Far-Field}} & 
        \text{\shortstack{Near-Field}} & 
        \text{\shortstack{Near-Field}} \\
        \hline 
        
        \textbf{Beam Scheme} & 
        Steering Beam (Gaussian) & 
        Focusing Beam (Gaussian) & 
        \text{Airy Beam} \\
        \hline 
        
        \textbf{Phase Profile} & 
        Linear Phase & 
        Quadratic Phase & 
        Cubic Phase \\
        \hline 

        \textbf{Parameters} & 
        $\theta$ & 
        $F,\theta$ & 
        $B,F,\theta$ \\
        \hline 
        
        \textbf{Expression} & 
        $\phi(x) = -k x \sin\theta$ & 
        $\phi(x) = -\frac{\pi}{\lambda F} x^2-kx\sin{\theta}$ & 
        $ \phi(x) = \frac{1}{3}(2\pi B)^3 x^3 - \frac{\pi}{\lambda F} x^2 - kx\sin{\theta} $ \\
        \hline 
        
        \textbf{LoS} & 
        % --- 关键修改：使用 margin 参数 ---
        % margin=左 下 右 上 (单位可以用pt或mm)
        % 这里设置上下各 10pt (约3.5mm) 的留白，彻底解决重合问题
        \includegraphics[width=0.25\linewidth, valign=c, margin=0pt 5pt 0pt 5pt]{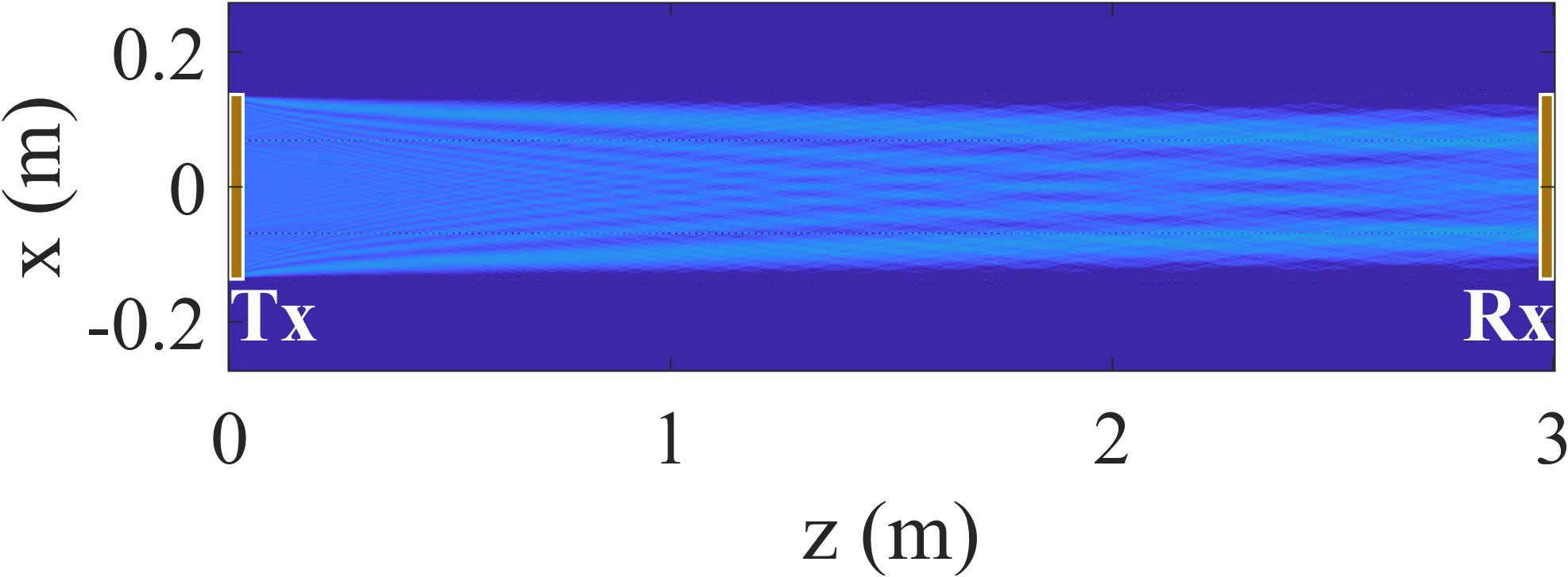} & 
        
        \includegraphics[width=0.25\linewidth, valign=c, margin=0pt 5pt 0pt 5pt]{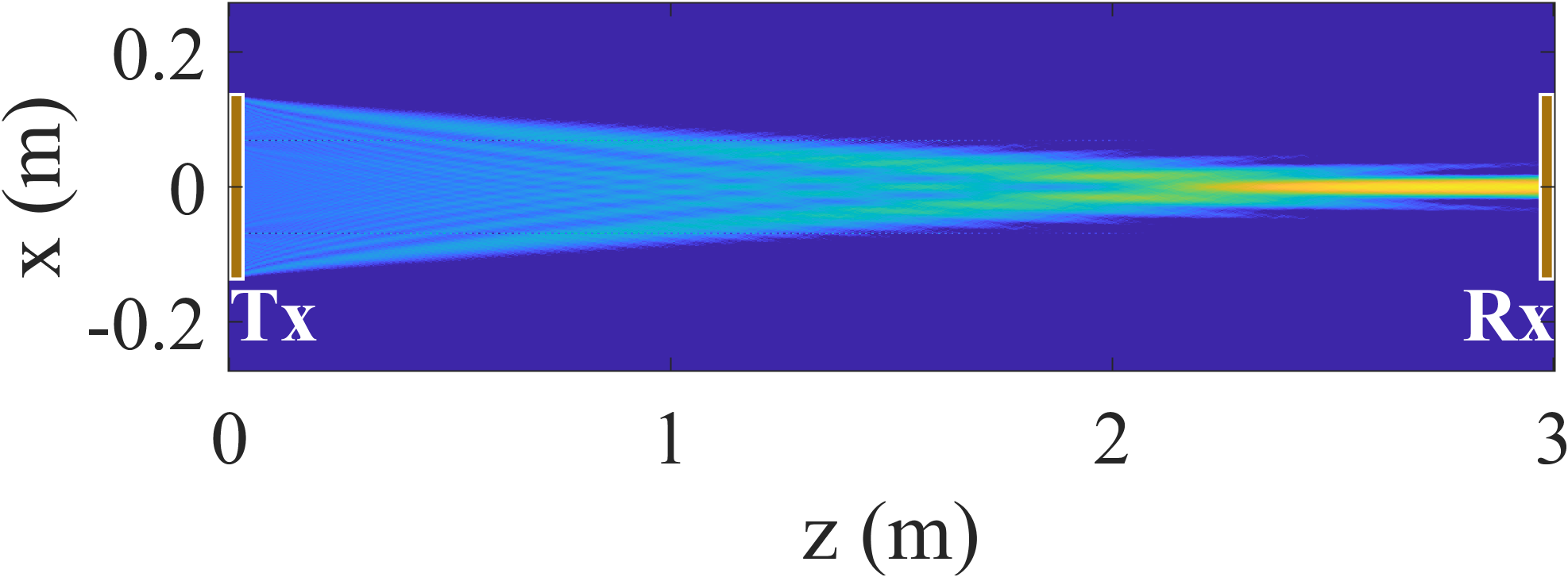} & 
        
        \includegraphics[width=0.25\linewidth, valign=c, margin=0pt 5pt 0pt 5pt]{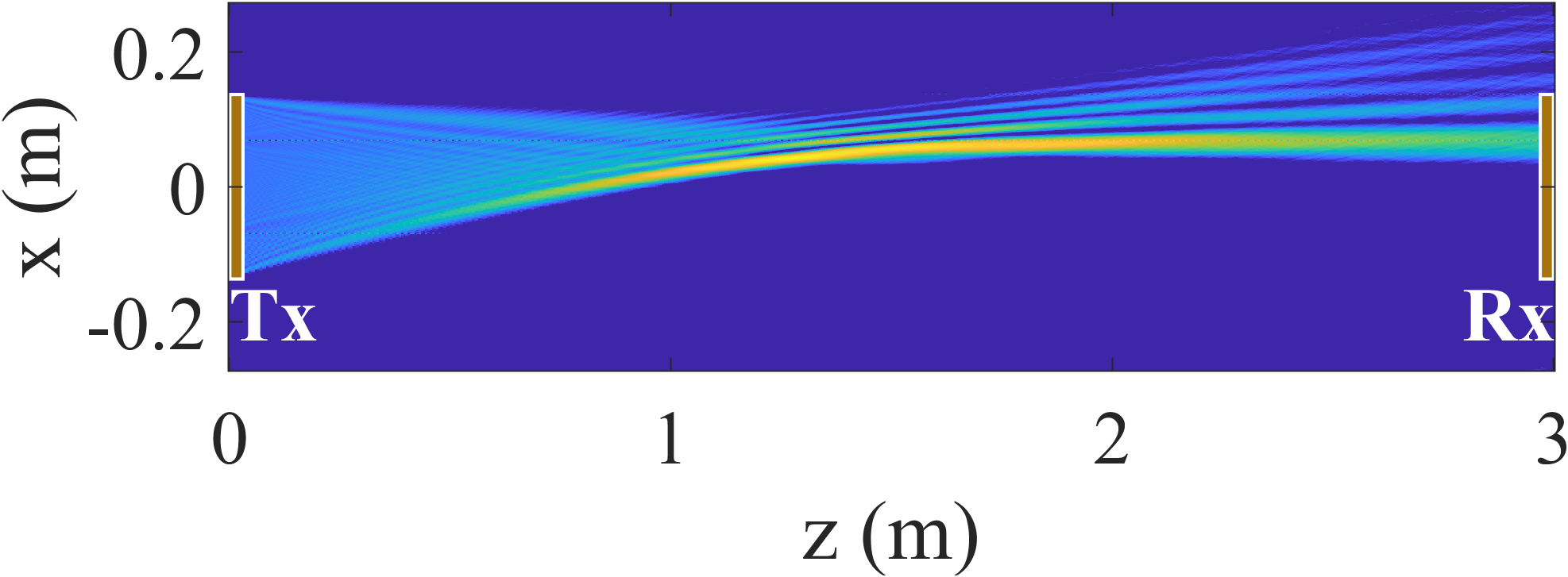} \\
        \hline 

         \textbf{Quasi-LoS} & 
        % --- 关键修改：使用 margin 参数 ---
        % margin=左 下 右 上 (单位可以用pt或mm)
        % 这里设置上下各 10pt (约3.5mm) 的留白，彻底解决重合问题
        \includegraphics[width=0.25\linewidth, valign=c, margin=0pt 5pt 0pt 5pt]{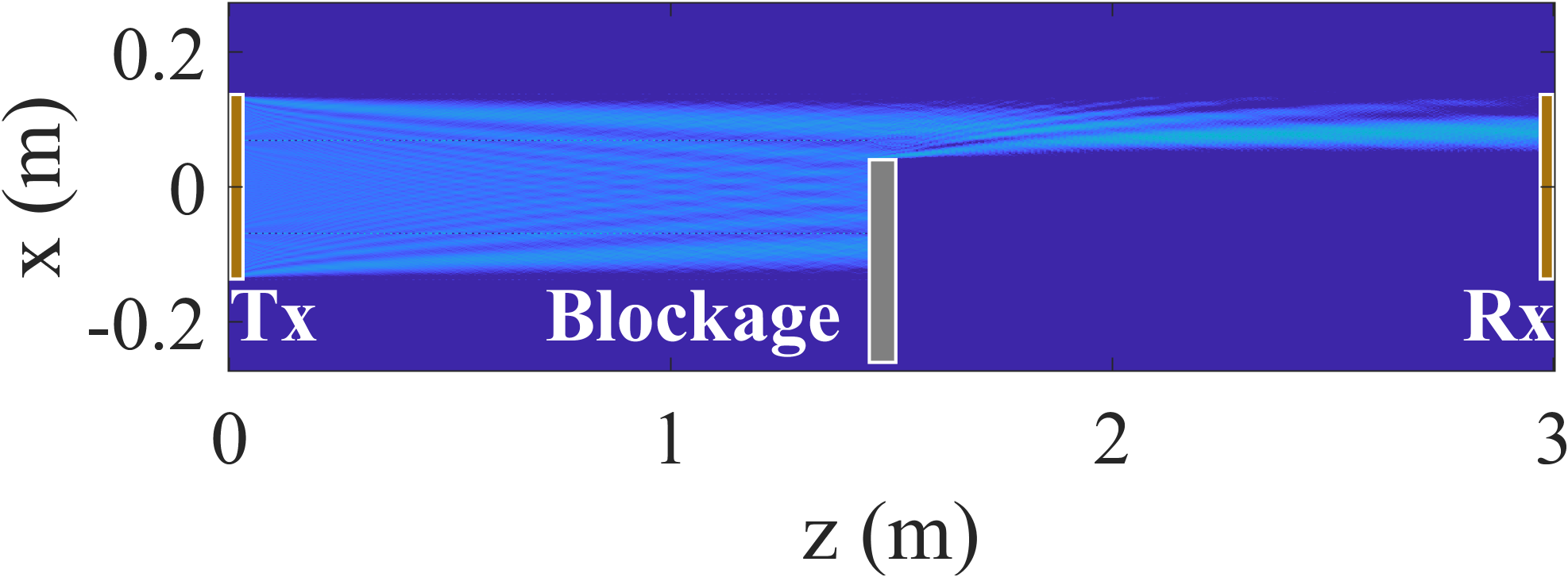} & 
        
        \includegraphics[width=0.25\linewidth, valign=c, margin=0pt 5pt 0pt 5pt]{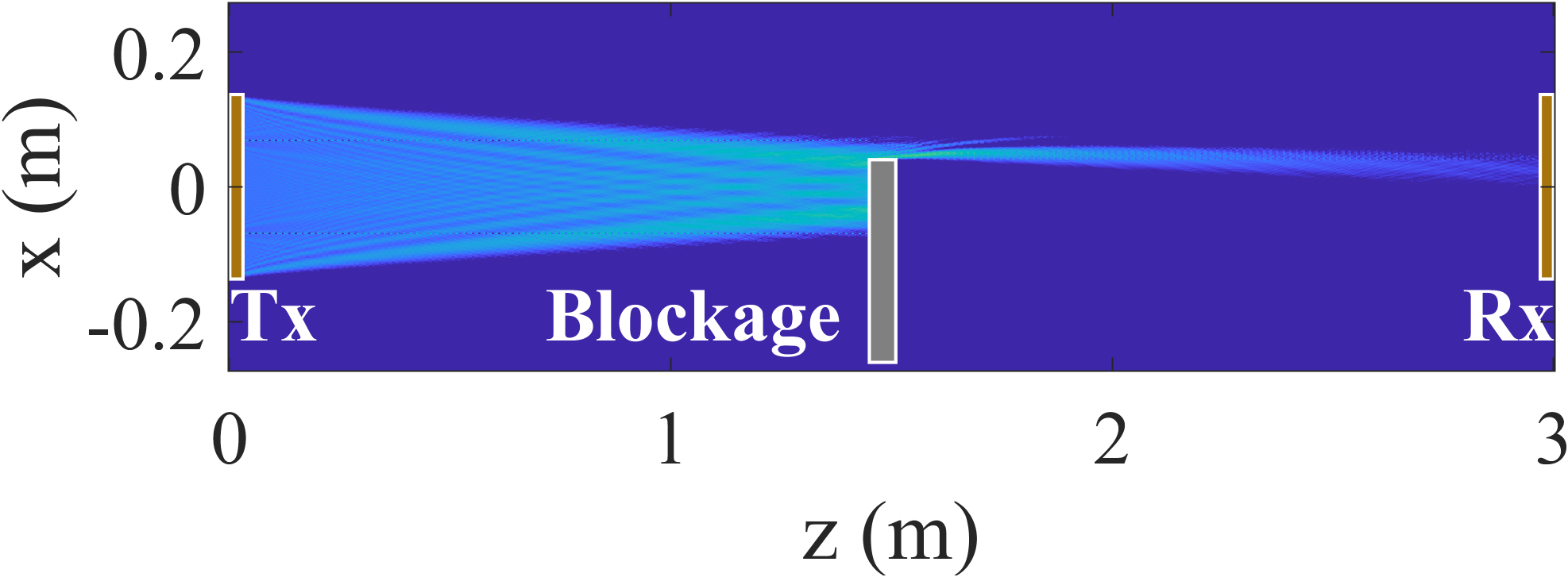} & 
        
        \includegraphics[width=0.25\linewidth, valign=c, margin=0pt 5pt 0pt 5pt]{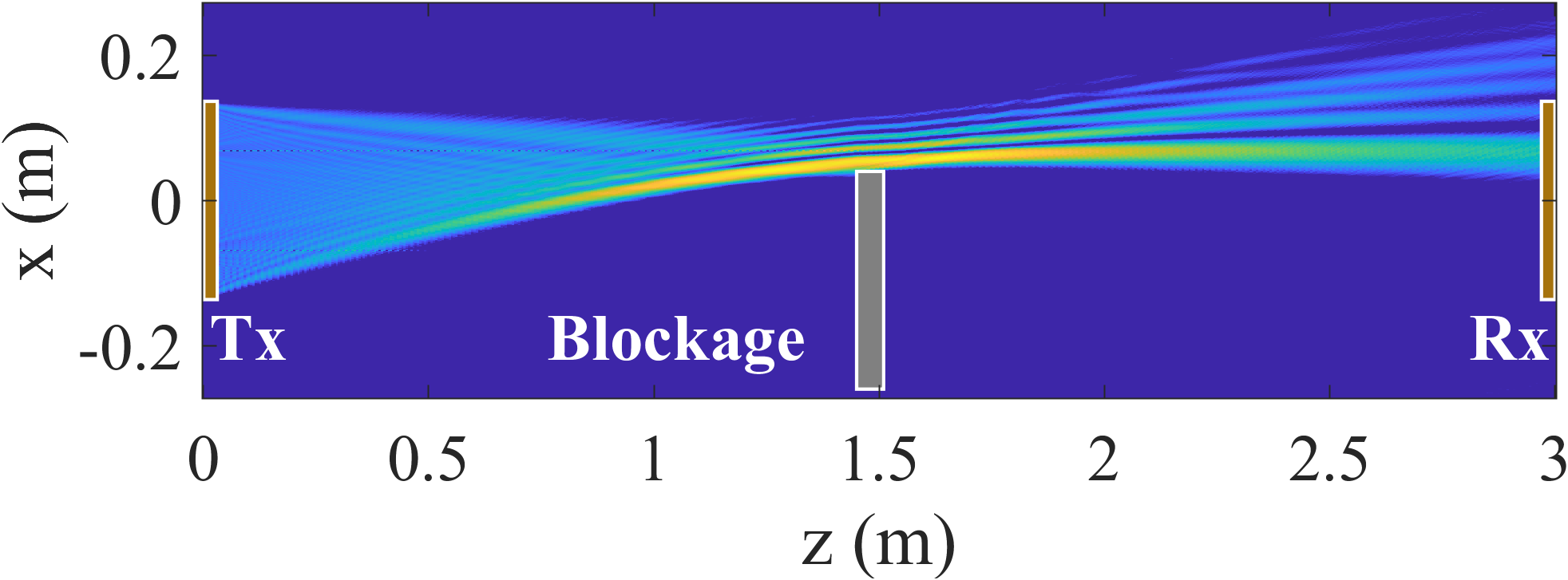} \\
        \hline 
    \end{tabular}
\end{table*}
% \begin{figure*}[t]
% \label{fig:intro}
%     \centering
%     \includegraphics[width=\linewidth]
%     {closed-form_Airy/images/Intro/Intro.png} % 使用 \linewidth 自适应栏宽
%     \caption{Comparison of propagation among focusing beam, SVD beamforming and Airy beam under LoS and quasi-LoS scenarios.}
% \end{figure*}
\subsection{Related Works}
The Airy beam, first proposed in the field of optics~\cite{Zhan-2020-Propagations,Nikolaos-2019-Airy},  has recently been proposed for use in the THz band~\cite{Jornet-2023-Wireless,Guerboukha-2024-Curving,Chen-2024-Curving,Chen-2025-physics,Yazdnian-2025-NirvaWave,Weng-2025-Learning,Zhao-2025-WDC,Petrov-2024-Wavefront,Petrov-2024-Wavefront-Hopping,Singh-2024-Wavefront}. In~\cite{Guerboukha-2024-Curving}, the authors conducted experimental measurements with a lens demonstrating that Airy beams carrying high-data-rate transmissions can establish links by bending around obstacles. In~\cite{Chen-2024-Curving}, the authors proposed a physics-informed learning-based framework to optimize the phase profile of the transmit array for generating Airy beams in a ULA system, which enables curved wavefront propagation to circumvent obstacles. This approach relies on a black-box neural network model with additional training process instead of a closed-form or interpretable analytical expression for the phase solution. Subsequently, the same team published a follow-up work in~\cite{Chen-2025-physics}, which experimentally validates the real-world feasibility of generating Airy beams with a metasurface platform. Their results further demonstrate that properly configured Airy beams can significantly improve the link budget in quasi-LoS scenarios, even outperforming near-field focused beams. 

For Airy beam training, the author in~\cite{Weng-2025-Learning} derived an approximation for the beam trajectory and propose a learning-based beam training method using the DFT codebook to identify the optimal configuration under unknown blockage conditions. For the practical applications, the authors in~\cite{Zhao-2025-WDC} first advocated the use of Airy beams in THz wireless data centers, and presented several low-complexity beam training schemes. To support the investigations of the Airy beam, an efficient near-field wave propagation simulator named NirvaWave~\cite{Yazdnian-2025-NirvaWave} has been developed, offering accurate and computationally efficient modeling of complex blocked environments to evaluate novel wavefronts like Airy beams. Furthermore, inspired by the curved trajectory of the Airy beam, the concept of wavefront hopping has been introduced to enhance physical-layer security and mitigate interference by dynamically switching beams over time~\cite{Petrov-2024-Wavefront,Petrov-2024-Wavefront-Hopping,Singh-2024-Wavefront}.

Although these studies have demonstrated the potential of Airy beams to bypass blockages and have proposed Airy beam design methods for quasi-LoS scenarios, most existing approaches remain confined to simple ULAs rather than more practical UPAs. Furthermore, they predominantly rely on deep learning-based methods that lack closed-form solutions, resulting in limited interpretability and constrained generalization in complex deployment environments.
 \subsection{Contributions}
To address the aforementioned challenges, in this paper, we first analyze the electric field, beam trajectory and field amplitude for Airy beam in ULA systems. Based on these analytical foundations, the closed-form Airy beam design for ULA is developed. Extending from ULA to more practical yet complicated UPA, closed-form solutions for UPA Airy beam design are derived based on the spatial separability of the 3D wavefront. Extensive simulation results demonstrate that the proposed closed-form Airy beam design in ULA and UPA outperforms conventional Gaussian beam in quasi-LoS scenarios while offering low computational complexity and enhanced physical interpretability. The main contributions are summarized as follows.
\begin{itemize}
    \item \textbf{We develop a closed-form Airy beam design method for ULA systems.} Specifically, we consider the ULA in the transceiver and derive the closed-form electric field distribution of the Airy beam using Fresnel diffraction integral. Consequently, the trajectory equation and electric field magnitude are derived. Then, based on these analysis, we give the Airy beam design method for ULA systems with closed-form mathematical expression of the three beam parameters, i.e., curving coefficient $B$, distance coefficient $F$ and angle coefficient $\theta$, which can be directly calculated once we know the position of the blockage and the transceiver. 

\item \textbf{We extend the Airy beam design in the UPA systems and derive the closed-form solutions with two operation modes.} With comprehensive analysis of the electric field, 3D beam trajectory and field amplitude, the Airy beam design in UPA systems is investigated which is decoupled into two sequential tasks. The first task is to identify the optimal bending dimension and determine the corresponding waypoint where the beam should bend over the blockage and received point to minimize path deviation. The second task is to synthesizing the Airy beam phase profile with two operation modes. First, the hybrid focusing-Airy mode combines a 1D closed-form Airy phase for obstacle circumvention with a standard focusing phase on the other dimension to concentrate energy at the receiver. Second, the dual Airy mode, applying independent closed-form Airy phase to both dimensions providing enhanced flexibility for complex blockage geometries. 

\item \textbf{We evaluate the performance of the proposed closed-form Airy beam design for both ULA and UPA systems compared with the conventional beamforming schemes.} We investigate the spectral efficiency and beam patterns with different beamforming schemes in different position of blockage. Extensive results demonstrate that the proposed closed-form design significantly outperforms conventional beamforming schemes in quasi-LoS scenarios. Furthermore, the proposed method achieves near-optimal performance relative to exhaustive search with significant reduced computational complexity and enhanced physical interpretability.

\end{itemize}

%\subsection{Organization}
The remainder of the paper is organized as follows. In Sec.~\ref{sec:system_model}, the system model and definition of blockage ratio for THz ULA and UPA UM-MIMO systems are provided. In Sec.~\ref{sec:ULA} we analyze the electric characteristics of Airy beam and develop the closed-form design in ULA system. Then the UPA Airy beam design is investigated in Sec.~\ref{sec:UPA}. After the numerical evaluation of the proposed closed-form Airy beam design methods in Sec.~\ref{sec:simulation}, the paper is summarized in Sec.~\ref{sec:conclusion}.

\section{System Model}
\label{sec:system_model}
In this section, we first introduce the considered THz communication systems in near field quasi-LoS scenarios with ULA and UPA, respectively. Next, the calculation of electromagnetic wave propagation is investigated, considering the physical blockage situations in the LoS region. This propagation model will serve as the foundation for the beam analysis presented in the following sections.

\subsection{Communication System}
We consider a point-to-point wireless communication system operating in the near-field THz band. The system is defined in the Cartesian coordinate system, where the signal propagates primarily along the positive $z$-axis. The transmitter (Tx) is located at the plane $z=0$, while the receiver (Rx) is placed parallel to the Tx at a propagation distance $z = D$.

We investigate 2D ULA and 3D UPA for the communication system.
In the 2D scenario, the Tx is equipped with a ULA composed of $N_t$ antennas aligned along the $x$-axis. The coordinate of the $n^{th}$ transmitting element, denoted as $(x_t^n,y_t^n,z_t^n)$, is given by:
\begin{equation}
(x_t^n,y_t^n,z_t^n) = \left( (n - \frac{N_t - 1}{2})d, \, 0, \, 0 \right), \quad n = 0, \dots, N_t-1,
\label{eq:ula_coords}
\end{equation}
where $d$ represents the antenna interval. For the general 3D scenario, the Tx employs a UPA on the $xy$-plane. The array consists of $N_t = N_x \times N_y$ elements, where $N_x$ and $N_y$ denote the number of elements along the $x$ and $y$ axes, respectively. The position of the $(m, n)$-th element is defined as
\begin{equation}
(x_t^{m,n},y_t^{m,n},z_t^{m,n})= \left( (m - \frac{N_x - 1}{2})d, \, (n - \frac{N_y - 1}{2})d, \, 0 \right),
\label{eq:upa_coords}
\end{equation}
where $m \in \{0, \dots, N_x-1\}$ and $n \in \{0, \dots, N_y-1\}$.
Similarly, the receiver is equipped with a ULA or UPA centered at $(0, 0, D)$. Its position $(x_r^n,y_r^n,z_r^n)$ or $(x_r^{m,n},y_r^{m,n},z_r^{m,n})$ is defined relative to the receiver center, satisfying the same geometric spacing $d$ as the transmitter.

\subsection{Electromagnetic Wave Propagation}

We adopt a channel modeling approach based on {scalar diffraction theory} and {Fourier principles} to accurately and efficiently simulate electromagnetic wave propagation in the {near-field} THz regime~\cite{Yazdnian-2025-NirvaWave}. Rather than the plane wave assumption adopted in the far-field, the foundation of this model is the {Rayleigh-Sommerfeld integral theory}~\cite{Chen-2024-Curving}, which is essential for capturing the {spherical wavefronts} required for accurate near-field analysis.
% II.B.1. Near-Field Wave Propagation Theory
% The fundamental physical principle governing wave evolution in free space is the Rayleigh-Sommerfeld integral theory. This theory posits that the electric field at the transmitter aperture can be viewed as an infinite array of point sources, each emitting spherical waves. The complex electric field E at any arbitrary observation point is then the superposition of the contributions from all these point sources.
% To bypass the computationally intensive nature of discrete integral calculation at every point in space , we utilize the Angular Spectrum Method (ASM). ASM simplifies the calculation by transforming the integral into the frequency domain via Fourier principles, interpreting free-space propagation as a linear system with a specific transfer function H(fx,fy)
To overcome the computational complexity of solving the integral, we utilize the {Angular Spectrum Method (ASM)}~\cite{Yazdnian-2025-NirvaWave}, leveraging the Discrete Fourier Transform. As a result, the electric field $E$ at the observation plane is calculated by taking the inverse Fourier transform of the product of the transformed initial field and the propagation transfer function $H(f_x, f_y)$. The propagation of the electric field $E$ is governed by
\begin{equation}
E(x',y',z'|E_0) = \mathcal{F}^{-1}\left\{ \mathcal{F}\{E(x,y,z_0)\} \times H(f_x, f_y) \right\},
\label{eq:asm_propagation}
\end{equation}
where $\mathcal{F}$ denotes the Fourier transform. The transfer function $H(f_x, f_y)$ in the spatial frequency domain is given by
\begin{equation}
H(f_x, f_y) = \exp\left(j 2\pi \frac{\delta_z}{\lambda} \sqrt{1 - \lambda^2 (f_x^2 + f_y^2)}\right),
\label{eq:transfer_function}
\end{equation}
where $\delta_z$ represents the propagation distance along the propagation axis, $\lambda$ is the wavelength, and $f_x, f_y$ are the spatial frequencies.

While standard ASM efficiently models free-space propagation, capturing the interaction with inhomogeneous mediums requires an iterative ASM scheme to model the diffraction caused by 3D physical obstacles\cite{Yazdnian-2025-NirvaWave}. We discretize the propagation path into steps of size $\delta_z$. A blockage mask $\mathrm{B_L}(x, y, z)$ is defined to characterize the 3D geometry of obstacles: $\mathrm{B_L}(x, y, z) = 1$ indicates free space, while $0 \le \mathrm{B_L}(x, y, z) = \alpha < 1$ represents the attenuation caused by blockage.
The electric field at the current propagation step $z$ is determined by applying the blockage mask to the field propagated from the previous step $z - \delta_z$, which can be expressed as
\begin{equation}
E(x, y, z) = \mathrm{B_L}\times \mathcal{F}^{-1} \left\{ H(f_x, f_y) \times \mathcal{F} \{ E(x, y, z - \delta_z) \} \right\}.
\end{equation}
This robust, iterative approach enables the efficient modeling of complex wave phenomena in the THz near-field environment with obstacles.

It is worth noting that the mathematical framework presented above is formulated for the general 3D scenario employing UPAs. For systems utilizing ULAs, this framework can be straightforwardly simplified to a {2D model}. This can be achieved by omitting the dependence on the vertical transverse dimension $y$ and replacing the 2D Fourier transforms with their 1D counterparts, thereby simplifying the computational domain to the $(x, z)$ plane.

\begin{figure}[t]
    \centering
    \subfigure[ULA.]{\includegraphics[width=0.6\linewidth]{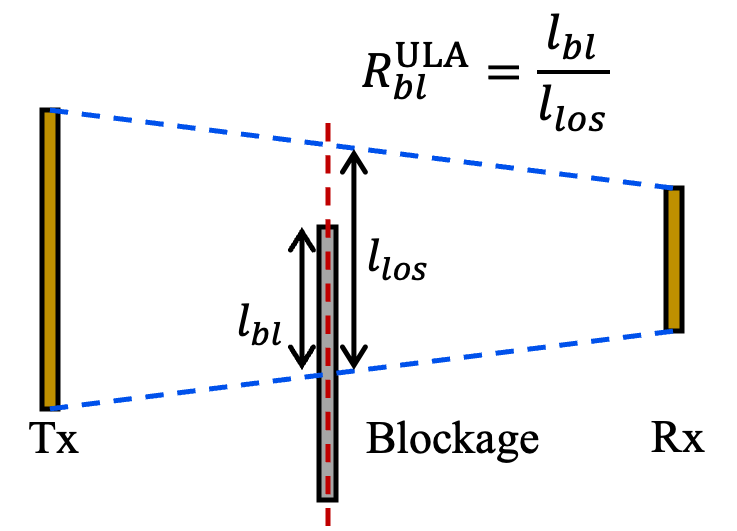}   \label{fig:Rate_ULA}}

     \subfigure[UPA.]{\includegraphics[width=0.7\linewidth]{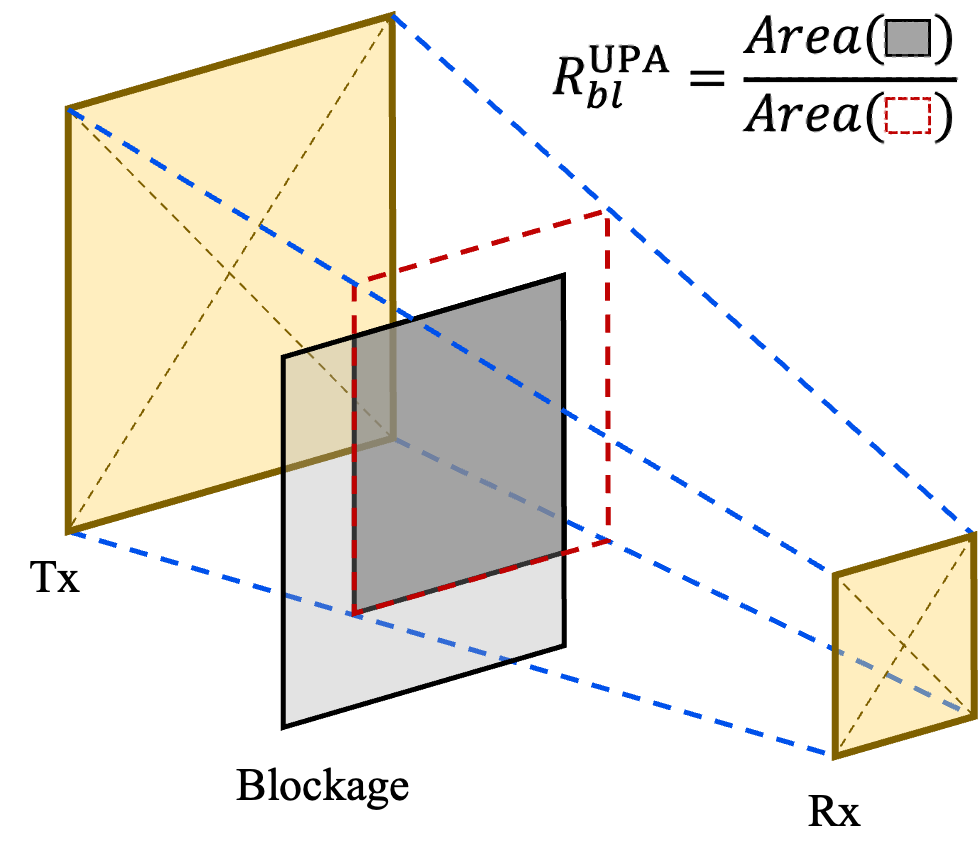} \label{fig:Rate_UPA}}
   \caption{Definition of the blockage ratio for ULA and UPA communication systems.}
   \label{fig:Rate}
\end{figure}

\subsection{Definition of the Blockage Ratio}
Given the finite physical apertures of the transmitter and receiver arrays, the spatial LoS propagation path forms a trapezoidal region for ULA or a frustum tunnel for UPA. To quantify the severity of blockage, we define a geometric blockage ratio at the location of obstacle. As illustrated in Fig.~\ref{fig:Rate}, for ULA systems, this ratio $R_{bl}^{{\textrm{ULA}}}$ is defined as the effective obstructed length $l_{bl}$ divided by the total vertical span of the LoS tunnel $l_{los}$. Similarly, for UPA systems, the ratio $R_{bl}^{{\textrm{UPA}}}$ is determined by the intersection area of the obstacle within the LoS tunnel relative to the total LoS area. This metric serves as a normalized measure of obstruction, where values of 0 and 1 represent LoS and NLoS conditions, respectively, while intermediate values denote quasi-LoS conditions.

\section{Closed-Form Airy Beam design for Uniform Linear Arrays}
% As we all know, for the near field beam focusing beam, once the position of the Tx and Rx is given, the phase profile of the focusing beam can be directly obtain according to the fix formulation \cite{Cui-2023-Near-Field-MIMO}. For the Airy beam, the 现有的 methods are learning-based methods~\cite{Chen-2024-Curving,Chen-2025-physics,Weng-2025-Learning}, where the environmental information of the Tx, Rx and blockage are input to the NN network to obtain the optimal configuration. To avoid the training process of the NN network and give an interpretable closed-form solution like beam focusing, in this section, we first consider the finite size of the transmit antenna array and analyze the electric field, main lobe trajectory and the electric amplitude. Then, an Airy beam design with closed-form solution for the phase profile under ULA is derived in both MISO and MIMO situation.
\label{sec:ULA}
% The phase profile of a near-field focusing beam can be directly derived from a closed-form expression once the positions of the Tx and Rx are specified~\cite{Cui-2023-Near-Field-MIMO}. In contrast, existing methods for configuring Airy beams predominantly rely on learning-based approaches~\cite{Chen-2024-Curving,Chen-2025-physics,Weng-2025-Learning}, where a neural network is trained to map environmental information including Tx and Rx positions and blockage geometry to an optimal beam configuration. 

% To avoid the need for NN training and to provide an interpretable, closed-form design analogous to that of focusing beams, 
In this section, we first account for the finite aperture of the transmit antenna array and conduct a thorough analysis of the electric field, beam trajectory, and field amplitude of the Airy beam. Building on this analysis, we subsequently derive a closed-form expression for Airy beam design method in ULA communication systems.

%上来一开始就要强调我们想要做的事情，就是像beam focusing一样，知道receiver位置我们就可以得到focusing beam的相位，那么我们也想推导，知道receiver和blockage的位置，我们也可以直接得到airy beam的相位，这样的方法低复杂度，而且不需要依赖神经网络。
%给出大公式，横跨整个页面的那种哈 电场表达式（18） E(x,z)
%blockage要给出不规则形状的哈
%UPA完事之后，然后再加上一个SUPA然后提出利用不规则消除旁瓣，然后进行仿真
\subsection{Electric Field}

According to the principle of Fourier optics~\cite{Nikolaos-2019-Airy}, the Airy beam can be generated by imposing a cubic phase to a Gaussian beam. The 1D phase profile is given by~\cite{Chen-2024-Curving,Chen-2025-physics}:
\begin{equation}
    \phi_x(x_0) = \frac{1}{3}(2\pi B)^3 x_0^3 - \frac{\pi}{\lambda F} x_0^2 - \frac{2\pi}{\lambda}\sin{\theta} x_0 ,
    \label{eq:airy_phase}
\end{equation}
where $\phi_x(x_0)$ denotes the phase distribution at the transmitter aperture, $x_0$ represents the transverse coordinate of the antenna array, and $\lambda$ is the wavelength of carrier frequency. The parameter $B$ governs the bending of the Airy beam, $F$ denotes the focal length, and $\theta$ represents the beam steering angle. Then, the initial electric field is $E(x_0,0)=e^{j\phi_x(x_0)}$. The 2D field propagation of the Airy beam with infinite array using the Fresnel diffraction integral with Fresnel approximation can be expressed as
    \begin{equation}
    E\left(x,z\right)=\frac{e^{jkz}}{j\lambda z}\int_{-\infty}^{+\infty}E\left(x_0,0\right)e^{j\frac{k}{2z}\left(x-x_0\right)^2}dx_0,
\end{equation}
where $k=\frac{2\pi}{\lambda}$. We denote the size of the aperture as $D_{arr}=(N_t-1)d$. Then the 2D field propagation of the finite Airy beam can be given by
\begin{equation}
    E\left(x,z\right)=\frac{e^{jkz}}{j\lambda z}\int_{-\infty}^{\infty}e^{j\phi_x\left(x_0\right)}e^{j\frac{k}{2z}\left(x-x_0\right)^2}\mathrm{rect}\left(\frac{x_0}{D_{arr}}\right)dx_0.
\end{equation}
While the $\mathrm{rect}$ function perfectly models the hard physical boundary of the array, the abrupt cut-off leads to sharp discontinuities in the spectrum. Specifically, the Fourier transform of the $\mathrm{rect}$ function, i.e., $\mathrm{sinc}$ function results in complex convolution integrals in the propagation equation, making the analytical derivation of the desired closed-form expression extremely difficult.
To ensure analytical tractability and simplify the derivation of closed-form solutions, we instead adopt a Gaussian window to model the finite aperture, which can be expressed as
\begin{equation}
  E(x,z)=\frac{e^{jkz}}{j\lambda z}\int_{-\infty}^{\infty}e^{j\phi_x(x_0)}e^{j\frac{k}{2z}(x-x_0)^2}e^{-\frac{x_0^2}{\omega_0^2}}dx_0,
\end{equation}
where $\omega_0$ represents the Gaussian beam waist and is set as $D_{arr}/2$. The primary advantage of using the Gaussian function is that its Fourier transform remains a Gaussian function, thereby preserving the mathematical form of the integral and enabling the derivation of the explicit closed-form solutions. By consolidating the terms inside the integral, the new initial field distribution $\widetilde{E}(x_0,0)$ can be concisely expressed as
       \begin{align}
           \widetilde{E}(x_0,0)&=e^{j\phi_x(x_0)}e^{-\frac{x_0^2}{\omega_0^2}} \nonumber\\
           % &=\exp{[j(\frac{1}{3}(2\pi B)^3x_0^3-\frac{\pi}{\lambda F}x_0^2-\frac{2\pi}{\lambda}\sin{\theta}x_0)-\frac{x_0^2}{\omega_0^2})]}\\
           &=\exp[j(\frac{(2\pi B)^3}{3}x_0^3-(\frac{\pi}{\lambda F}-\frac{j}{\omega_0^2})x_0^2-\frac{2\pi}{\lambda}\sin{\theta}x_0)],
       \end{align}
   % \begin{subequations}
   %      \begin{align}
   %          \widetilde{E}(x_0,0)&=e^{j\phi_x(x_0)}e^{-\frac{x_0^2}{\omega_0^2}}\\
   %          \label{eq:jj}
   %          &=\exp\left[j\left(\frac{1}{3}(2\pi B)^3x_0^3-\left(\frac{\pi}{\lambda F}-\frac{j}{\omega_0^2}\right)x_0^2-\frac{2\pi}{\lambda}\sin{\theta}x_0\right)\right]
   %      \end{align}
   %  \end{subequations}
where the last equality follows from $-\frac{x_0^2}{w_0^2}=j^2\frac{x_0^2}{w_0^2}$. We denote $\widetilde{F}$ as the complex focusing distance satisfying $\frac{\pi}{\lambda F}-\frac{j}{\omega_0^2}=\frac{\pi}{\lambda \widetilde{F}}$. The equation can be transformed into 
\begin{equation}
    \frac{1}{F}-j\frac{\lambda}{\pi w_0^2}=\frac{1}{\widetilde{F}}.
\end{equation}
Then, we plug the initial electric field $\widetilde{E}(x_0,0)$ into the Rayleigh-Sommerfeld propagation integral yields the field $E(x, z)$ at distance $z$, which can be expressed as
    % \begin{subequations}
    % \begin{align}
    %      E(x,z)&=\frac{e^{jkz}}{j\lambda z}\int_{-\infty}^{+\infty}\exp (j(\frac{(2\pi B)^3}{3}x_0^3-(\frac{\pi}{\lambda \widetilde{F}})x_0^2 \nonumber\\ &-\frac{2\pi}{\lambda}\sin{\theta}x_0+\frac{\pi}{z}(x-x_0)^2)dx_0)\\ 
    %      &=\frac{e^{jkz}}{j\lambda z}\int_{-\infty}^{+\infty} \exp (j(\frac{(2\pi B)^3}{3}x_0^3+(\frac{\pi}{\lambda z}-\frac{\pi}{\lambda \widetilde{F}})x_0^2\nonumber\\&-(\frac{2\pi}{\lambda}\sin{\theta}+\frac{2\pi}{\lambda z}x)x_0+\frac{\pi}{\lambda z}x^2)) dx_0
    % \end{align}
    % \end{subequations}
\begin{align}
\label{eq:propagation_integral_split}
    E(x,z)&=\frac{e^{jkz}}{j\lambda z}\int_{-\infty}^{+\infty} \exp \left[ j \left( \frac{(2\pi B)^3}{3}x_0^3+\left(\frac{\pi}{\lambda z}-\frac{\pi}{\lambda \widetilde{F}}\right)x_0^2 \right. \right. \nonumber\\
    &\quad \left.\left. - \left(\frac{2\pi}{\lambda}\sin{\theta}+\frac{2\pi x}{\lambda z}\right)x_0+\frac{\pi x^2}{\lambda z} \right) \right] dx_0.
\end{align}
    %We focusing on the exponential term in the integrand.
    % \begin{subequations}
    %     \begin{align}
    %         &\frac{(2\pi B)^3}{3}x_0^3-\frac{\pi}{\lambda \widetilde{F}}x_0^2-\frac{2\pi}{\lambda}\sin{\theta}x_0+\frac{k}{2z}x^2-\frac{k}{z}xx_0+\frac{k}{2z}x_0^2\\
    %         &\frac{(2\pi B)^3}{3}x_0^3+(\frac{\pi}{\lambda z}-\frac{\pi}{\lambda \widetilde{F}})x_0^2-(\frac{2\pi}{\lambda}\sin{\theta}+\frac{2\pi}{\lambda z}x)x_0+\frac{\pi}{\lambda z}x^2
    %     \end{align}
    % \end{subequations}
We denote $A=(2\pi B)^3$, $C_1=-(\frac{2\pi}{\lambda}\sin{\theta}+\frac{2\pi}{\lambda z}x)$ and $C_2=\frac{\pi}{\lambda}(\frac{1}{z}-\frac{1}{\widetilde{F}})$ and the electric field can be concisely expressed as 
    \begin{equation}
    \label{eq:need_appendix}
        E(x,z)=\frac{e^{j(kz+\frac{\pi x^2}{\lambda z})}}{j\lambda z}\int_{-\infty}^{\infty}\exp [{j(\frac{A}{3}x_0^3+C_2x_0^2+C_1x_0)}]dx_0.
    \end{equation}
The final analytical closed-form of the electric field $E(x,z)$ is obtained by eliminating the quadratic term $C_2 x_0^2$ via a proper coordinate transformation and utilizing the standard definition of the Airy function $\mathbf{Ai}(\cdot)$, which is presented as
\begin{equation}
    \int_{-\infty}^{\infty}\exp\left[j\left(\frac{\tau^3}{3}+\xi\tau\right)\right]d\tau=2\pi \mathbf{Ai}(\xi).
\end{equation}
Then the closed-form expression for the electric field $E(x,z)$ of the Airy beam generated by the ULA under Gaussian window and the Fresnel approximation is presented by  
% \begin{equation}
%         E(x,z)=\frac{e^{jkz}}{j\lambda z B}e^{j\frac{\pi}{\lambda z}x^2}e^{j(\frac{2C_2^3}{3A^2}-\frac{C_1C_2}{A})}\mathbf{Ai}(\xi)
%     \end{equation}
\begin{equation}
    E(x,z)=\frac{e^{jkz}}{j\lambda z B}e^{j\frac{\pi}{\lambda z}x^2} \exp \left[ j \left( \frac{2C_2^3}{3A^2}-\frac{C_1C_2}{A} \right) \right] \mathbf{Ai}(\xi),
\label{eq:electric_field}
\end{equation}
where $\xi$ is the independent variable of the Airy function and can be expressed as 
\begin{equation}
\xi= -\frac{\sin{\theta}}{\lambda B}-\frac{1}{\lambda z B}x-\frac{\left(\frac{1}{z}-\frac{1}{\widetilde{F}}\right)^2}{16\lambda^2\pi^2B^4}.
\end{equation}
The complete derivation of the closed-form expression \eqref{eq:electric_field} of the electric field is provided in \textbf{Appendix}~\ref{appendix:A}.
\subsection{Trajectory}

\begin{figure}[t]
    \centering
    \includegraphics[width=0.9\linewidth]{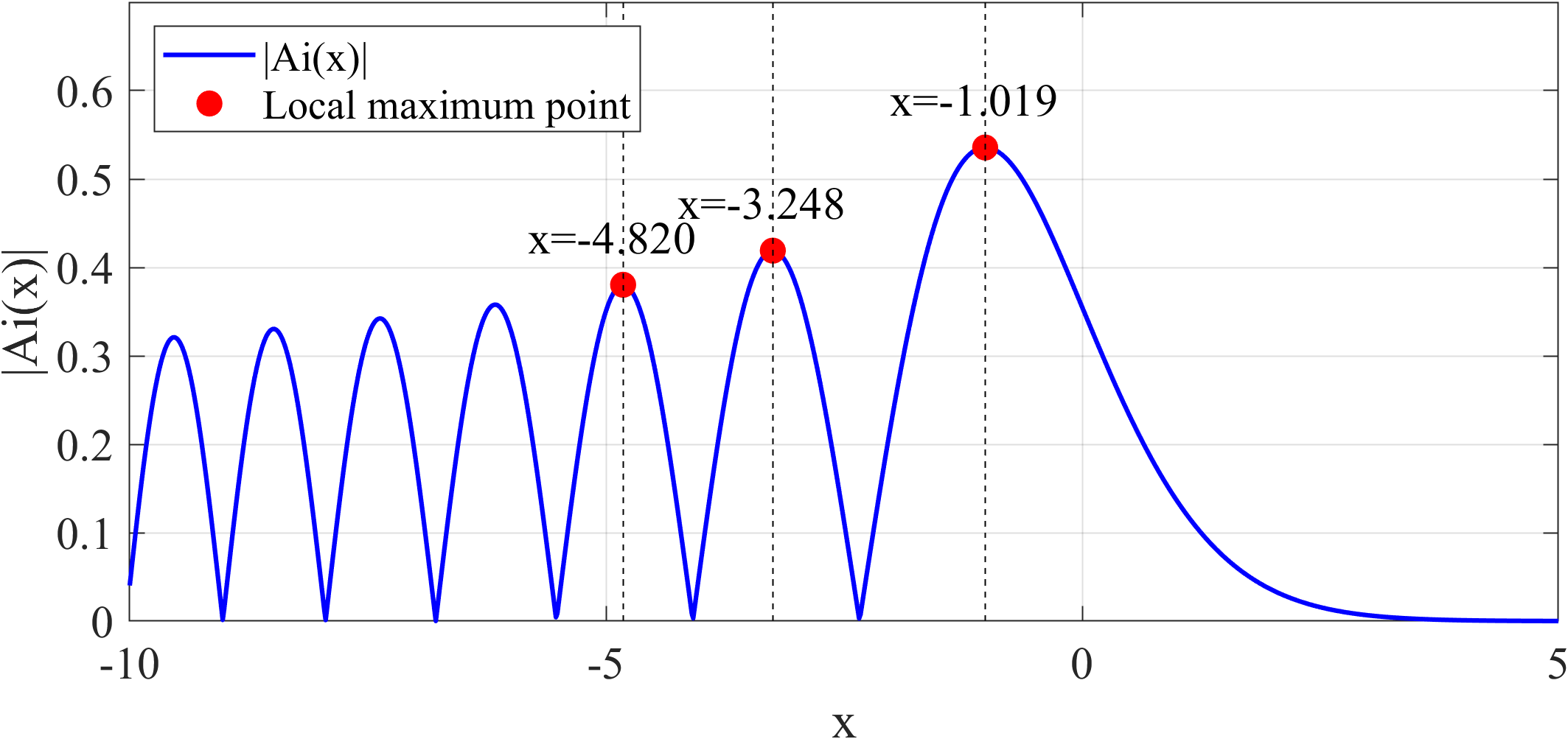} % 使用 \linewidth 自适应栏宽
    \caption{Amplitude and the local maximum points of the Airy function $\mathbf{Ai(x)}$.}
\label{fig:airy_function}
\end{figure}
\begin{figure}[t]
    \centering
    \includegraphics[width=0.9\linewidth]{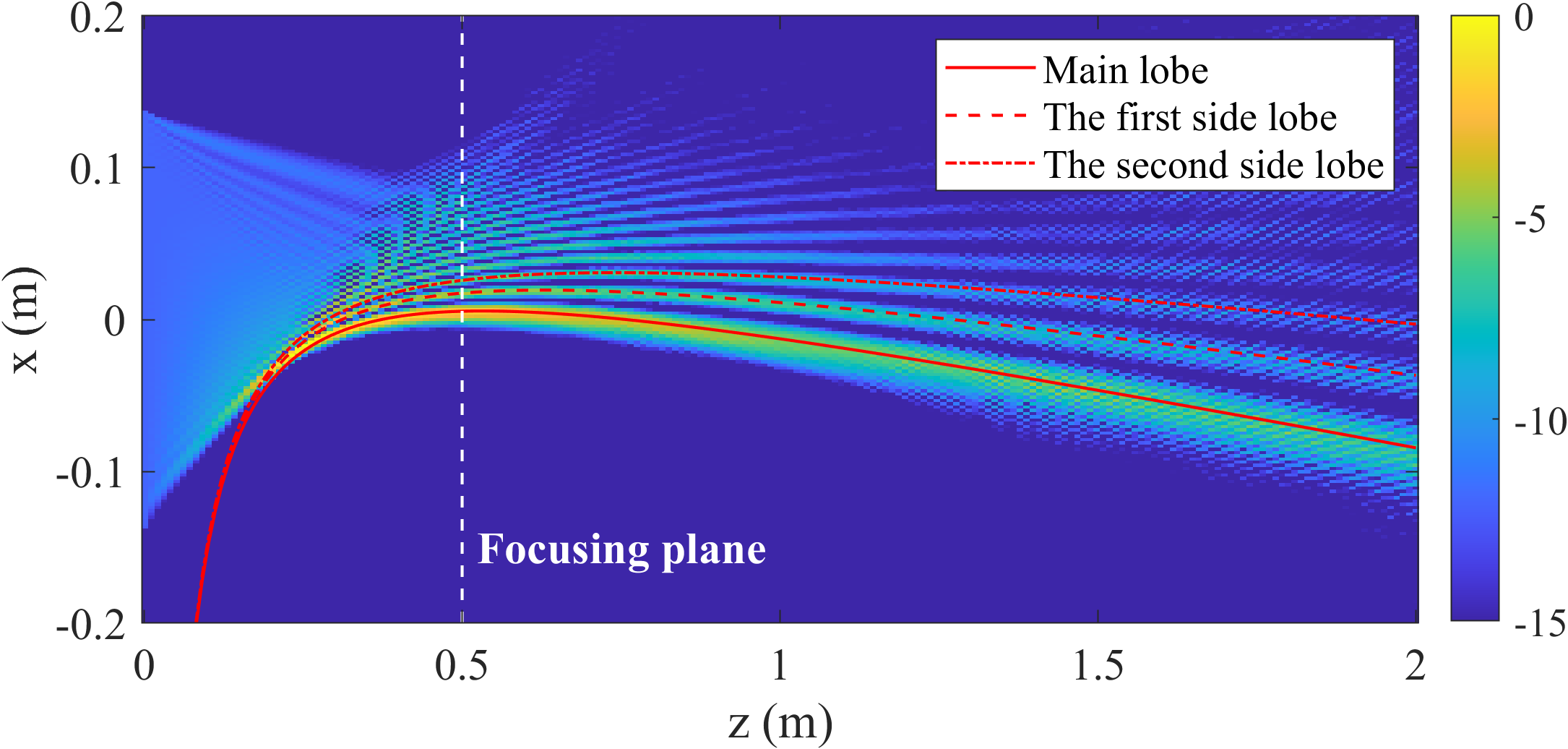} % 使用 \linewidth 自适应栏宽
    \caption{Validation of the closed-form trajectory equation. ($N_t=256$, $D_{arr}=0.2732$ m, $d=\lambda/2$ at $140\text{ GHz}$, $B=5$,$F=0.5$, $\theta=-0.03$).}
\label{fig:tra}
\end{figure}

According to \eqref{eq:electric_field}, the expression contains an Airy function, implying that the main and side lobes are located where $\xi$ is the local maximum point of the amplitude of the Airy function $|\mathbf{Ai(\cdot)}|$. As shown in Fig.~\ref{fig:airy_function}, to determine the trajectory of the main lobe, we set the real part of $\xi$ equal to $\xi_{peak}\approx-1.0188$, which corresponds to the first local maximum point of the Airy function.
Let $S=\frac{1}{z}-\frac{1}{\widetilde{F}}=(\frac{1}{z}-\frac{1}{F})+j(\frac{\lambda}{\pi \omega_0^2})$ with its real and imaginary parts denoted as $S_R=\frac{1}{z}-\frac{1}{F}$ and $S_I=\frac{\lambda}{\pi\omega_0^2}$, respectively. Hence, $\mathcal{R}\{S^2\}=S_R^2-S_I^2$. Substituting these into the real part of $\xi$, which can be given by
    \begin{equation}
        \mathcal{R}\{\xi\}=-\frac{\sin{\theta}}{\lambda B}-\frac{1}{\lambda z B}x-\frac{S_R^2-S_I^2}{16\lambda^2\pi^2B^4}=\xi_{peak}.
    \end{equation}
Next, we can derive the closed-form trajectory of the main lobe, which can be expressed as   
    \begin{equation}
    \label{eq:trajectory}
        x(z)=-\xi_{peak}\lambda z B-\sin{\theta}z-\frac{S_R^2-S_I^2}{16\lambda\pi^2B^3}z.
    \end{equation}
Similarly, by setting $\xi$ to the second and third maxima of the Airy function equal to $-3.248$ and $-4.820$, we can obtain the trajectories of the first and second side lobes, respectively. The validation of the closed-form trajectory equation is shown in Fig.~\ref{fig:tra}.

\begin{figure}[t]
    \centering
    \includegraphics[width=0.9\linewidth]{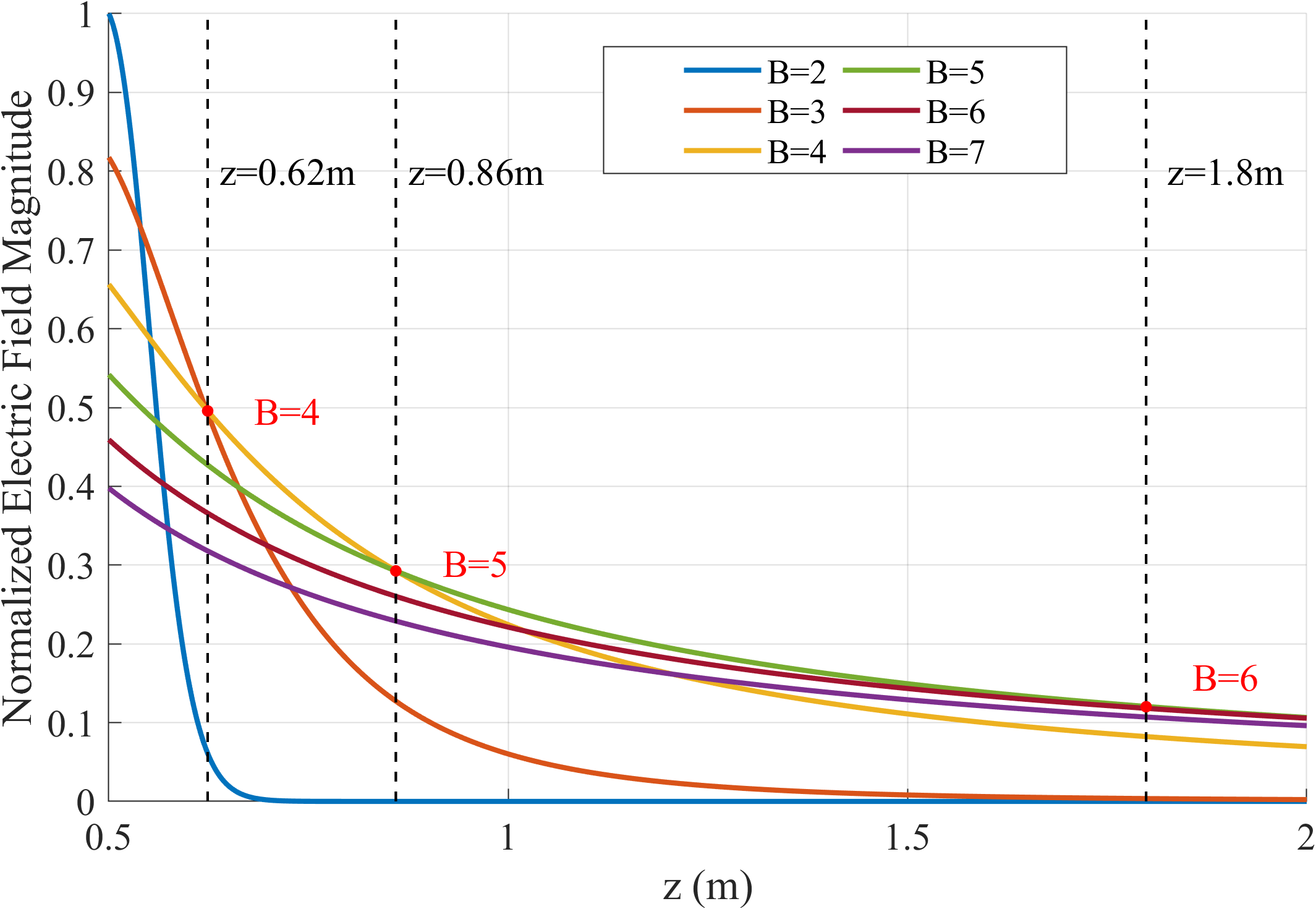} % 使用 \linewidth 自适应栏宽
    \caption{Electric field magnitude on the main lobe trajectory with different $B$. ($N_t=256$, $D_{arr}=0.2732$ m, $d=\lambda/2$ at $140\text{ GHz}$, $F=0.5\text{ m}$).}
\label{fig:magnitude}
\end{figure}

\subsection{Electric Field Magnitude}
% According to \textit{Lemma 1}, we can continuously calculate the electric field magnitude, which can be expressed as 
% \begin{equation}
%     |E(x,z)|=|\frac{1}{\lambda z B}||\exp{[j(\frac{2C_2^3}{3A^2}-\frac{C_1C_2}{A})]}||\mathbf{Ai}(\xi)|
% \end{equation}
% Since we focus on the electric amplitude of the main lobe, $|\mathbf{Ai}(\xi)|$ always a constant denoted as $C_{Ai}$, where $\xi$ satisfies $\mathcal{Re}\{\xi\}=\xi_{peak}$. According to $|e^{j\phi_{c}}|=e^{-\mathit{Im}(\phi_c)}$, we denote $\phi_{c}=\frac{2C_2^3}{3A^2}-\frac{C_1C_2}{A}$. $C_1$is real and $C_2$ is complex. Follow the trajectory equation, $\xi_{peak}=\frac{1}{A^{1/3}}(C_1-\frac{\mathrm{Re}(C_2^2)}{A})$. Then we can get $C_1=\xi_{peak}A^{\frac{1}{3}}+\frac{\mathrm{Re}(C_2^2)}{A}$. plugging into $\phi_c$
%     \begin{subequations}
%     \begin{align}
%              \phi_c&=-\xi_{peak}\frac{C_2}{A^{\frac{2}{3}}}+\frac{1}{A^2}(\frac{2}{3}C_2^3-C_2\mathrm{Re}(C_2^2))\\
%              &=-\xi_{peak}\frac{C_2}{(2\pi B)^2}+\frac{1}{(2\pi B)^6}(\frac{2}{3}C_2^3-C_2\mathrm{Re}(C_2^2))
%     \end{align}
%     \end{subequations}
Based on~\eqref{eq:electric_field}, the magnitude of the electric field can be continuously evaluated as
\begin{equation}
    |E(x,z)| = \left| \frac{1}{\lambda z B} \right| \cdot \left| \exp(j \phi_c) \right| \cdot |\mathbf{Ai}(\xi)|,
\end{equation}
where the complex phase term is defined as $\phi_c = \frac{2C_2^3}{3A^2} - \frac{C_1 C_2}{A}$. Since our analysis focuses on the peak intensity of the main lobe, $|\mathbf{Ai}(\xi)|$ is treated as a constant denoted by $C_{\mathrm{Ai}}$, provided that $\mathrm{Re}\{\xi\} = \xi_{\mathrm{peak}}$. 
Given that $C_1$ is a real-valued constant and $C_2$ is complex, the magnitude of the exponential term is determined by the imaginary part of its argument, i.e., $|\exp(j \phi_c)| = \exp(-\mathrm{Im}\{\phi_c\})$. According to the trajectory equation \eqref{eq:trajectory}, the peak condition is satisfied when $C_1 = \xi_{\mathrm{peak}} A^{1/3} + \frac{\mathrm{Re}\{C_2^2\}}{A}$. Substituting this into the expression for $\phi_c$, it can be expressed as 
\begin{align}
\label{eq:phi_c_derivation}
    \phi_c &= -\xi_{\mathrm{peak}} \frac{C_2}{A^{2/3}} + \frac{1}{A^2} \left( \frac{2}{3}C_2^3 - C_2 \mathrm{Re}\{C_2^2\} \right) \nonumber\\
    &= -\xi_{\mathrm{peak}} \frac{C_2}{(2\pi B)^2} + \frac{1}{(2\pi B)^6} \left( \frac{2}{3}C_2^3 - C_2 \mathrm{Re}\{C_2^2\} \right).
\end{align}
%    abstract the imagine part
%    \begin{subequations}
%    \begin{align}
%         K_2&=-\frac{\xi_{peak}\mathrm{Im}(C_2)}{(2\pi)^2}=-\frac{\xi_{peak}I_{C_2}}{(2\pi)^2},\\K_6&=\frac{\mathrm{Im}(\frac{2}{3}C_2^3-C_2\mathrm{Re}(C_2^2))}{(2\pi)^6}=\frac{R_{C_2}^2I_{C_2}+\frac{1}{3}I_{C_2}^3}{(2\pi)^6}\\
%         \mathrm{Im}({\phi_c})&=\frac{K_2}{B^2}+\frac{K_6}{B^6}
%    \end{align}
%    \end{subequations}
%    where $R_{C_2}=\mathrm{Re}(C_2)$ and  $I_{C_2}=\mathrm{Im}(C_2)$
% Then
% \begin{equation}
% \label{eq:electric_amplitude}
% |E(B)|=|\frac{1}{\lambda z B}||\exp{-(\frac{K_2}{B^2}+\frac{K_6}{B^6})}|C_{Ai}
% \end{equation}
Let $R_{C_2} = \mathrm{Re}\{C_2\}$ and $I_{C_2} = \mathrm{Im}\{C_2\}$ denote the real and imaginary parts of $C_2$, respectively. The imaginary part of $\phi_c$ can then be expressed as 

\begin{equation}
      \mathrm{Im}\{\phi_c\} = \frac{K_2}{B^2} + \frac{K_6}{B^6},
\end{equation}
where $K_2$ and $K_6$ are the coefficient of $1/B^2$ and $1/B^6$ which are defined as
\begin{subequations}
\label{eq:Im_phi_c}
\begin{align}
    K_2 &= -\frac{\xi_{\mathrm{peak}} I_{C_2}}{(2\pi)^2}, \\
    K_6 &= \frac{\mathrm{Im}\left\{ \frac{2}{3}C_2^3 - C_2 \mathrm{Re}\{C_2^2\} \right\}}{(2\pi)^6} = \frac{R_{C_2}^2 I_{C_2} + \frac{1}{3} I_{C_2}^3}{(2\pi)^6}.
\end{align}
\end{subequations}
Consequently, substituting \eqref{eq:Im_phi_c} into the magnitude expression, the final analytical form of the electric field amplitude as a function of the parameter $B$ and $F$ is given by
\begin{equation}
\label{eq:electric_amplitude}
\left|E(B,F)\right| = \left| \frac{C_{\mathrm{Ai}}}{\lambda z B} \right| \exp \left[ -\left( \frac{K_2}{B^2} + \frac{K_6}{B^6} \right) \right].
\end{equation}

Fig. \ref{fig:magnitude} depicts the electric field magnitude along the propagation trajectory with different $B$ values. We observe that with the fixed $F$, an increase in $B$ leads to a reduction in magnitude at the focusing plane but simultaneously yields a slower rate of attenuation. This occurs because a larger $B$ distributes more energy to the side lobes, lowering the initial intensity of the main lobe. However, as the wave propagates, these side lobes supply more energy to the main lobe, resulting in a more sustained electric field magnitude.
Owing to this phenomenon, the optimal Airy beam configuration characterized by the parameter $B$ is inherently distance-dependent. This behavior is demonstrated in Fig. \ref{fig:magnitude}, which compares the electric field magnitude of six values of the bending parameter B. For instance, within the short-to-medium range of $ z\in [0.62,0.86]$ m, the beam with $B=4$ achieves the highest energy. However, as the propagation distance increases to the interval $z\in[0.86,1.8]$ m, the configuration with $B=5$ becomes superior. Beyond $z=1.8$ m, the beam with $B=6$ eventually outperforms the others due to its enhanced energy supplement from the side lobes. 
% These results underscore the necessity of adaptively optimizing the Airy beam parameters based on the specific location of the receiver to maximize the received signal.
% ln
% \begin{equation}
%   L(B)= \ln|E(B)|=\ln{C_{Ai}}-\ln{\lambda z}-\ln{B}-\mathrm{Im}({\phi_c})
% \end{equation}
% Differentiate the expression with respect to B
% \begin{equation}
% \frac{dL(B)}{dB}=-\frac{1}{B}+\frac{2K_2}{B^3}+\frac{6K_6}{B^7}
% \end{equation}
% \begin{equation}
%     B^6-2K_2B^4-6K_6=0
% \end{equation}
% To get the solution and select the current solution
% let $T=y-\frac{2K_2}{3}$, transform into standard form
% \begin{equation}
%     y^3+py+q=0
% \end{equation}
% where $p=-\frac{4K_2^2}{3}$,$q=-\frac{16K_2^3}{27}-6K_6$ 
% calculate $\Delta=(\frac{q}{2})^2+(\frac{p}{3})^3$
% obtain $y$
% \begin{equation}
% \label{eq:cal_B}
%     y=\sqrt[3]{-\frac{q}{2}+\sqrt{\Delta}}+\sqrt[3]{-\frac{q}{2}-\sqrt{\Delta}}
% \end{equation}
% $T=y+\frac{2K_2}{3}$ and $B=\sqrt{T}$

% \begin{figure}[ht]
%     \centering
%     \subfigure[Geometrical channel model.]{\includegraphics[width=\linewidth]{closed-form_Airy/images/figure_test/focusing_case1.png}\label{fig:GCM}} 
%      \subfigure[Wave-based channel model.]{\includegraphics[width=\linewidth]{closed-form_Airy/images/figure_test/airy_case1.png}\label{fig:WCM}}
%      \subfigure[Cascaded geometric and wave channel model. ]{\includegraphics[width=\linewidth]{closed-form_Airy/images/figure_test/beam_comp.png}\label{fig:CGWCM}}
%    \caption{The accuracy of the GCM and CGWCM.}
% \end{figure}

\subsection{Airy Beam Design in ULA}

\begin{figure}[t]
    \centering
    \includegraphics[width=0.8\linewidth]{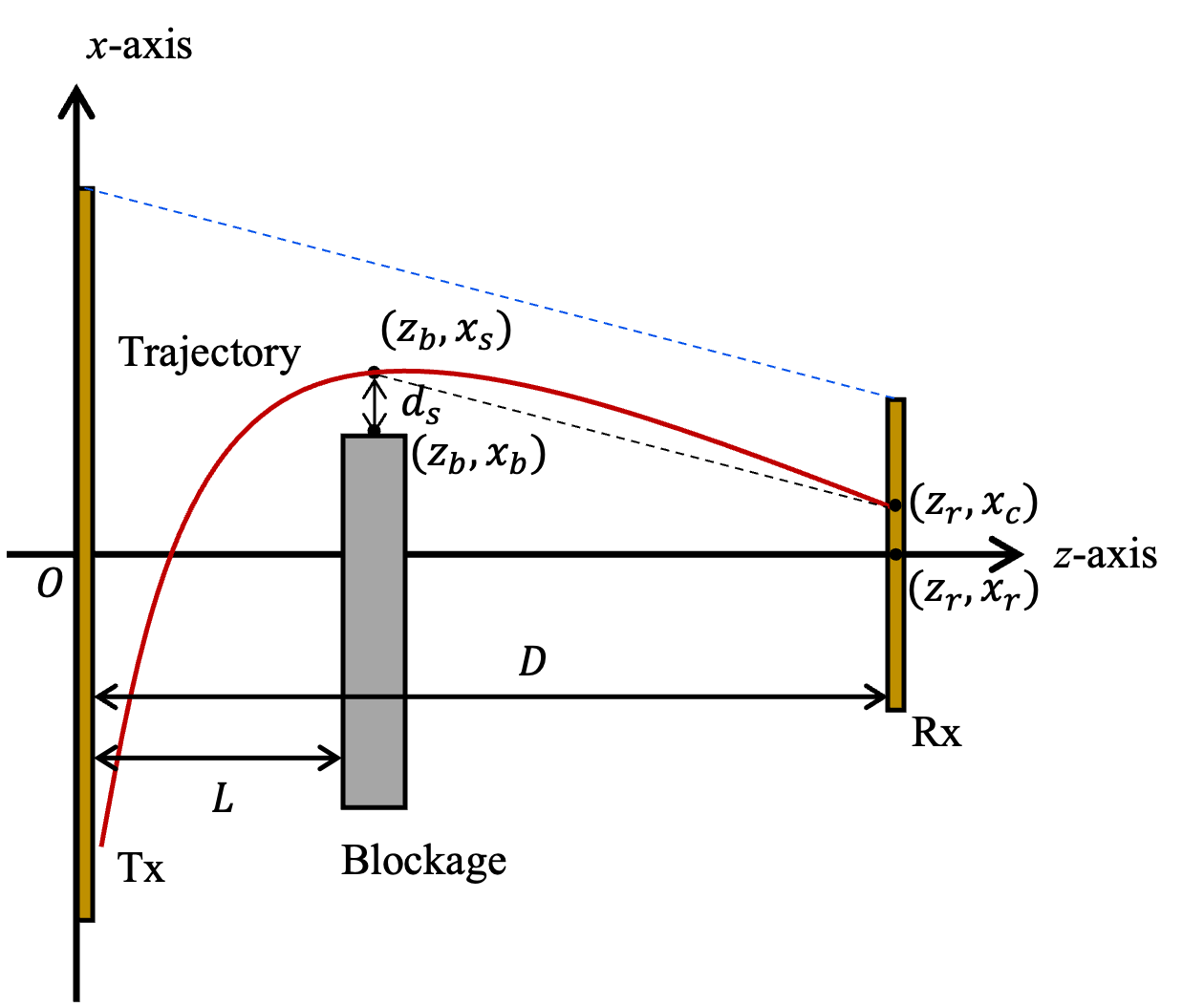} % 使用 \linewidth 自适应栏宽
    \caption{Airy beam design in ULA communication systems.}
\label{fig:ULA_airy_Design}
\end{figure}

Based on the preceding analytical derivations of the propagation trajectory and field magnitude, we now formulate the unified design framework for the Airy beam in ULA systems. The primary objective is to identify the optimal set of parameters $\{B, F, \theta\}$ that maximizes the received electric field magnitude at the receiver array, while ensuring the beam trajectory effectively bends over the blockage. To achieve a computationally efficient and interpretable solution, we decouple the design problem into two sequential tasks, i.e., determining the geometric trajectory constraints and deriving the closed-form parameters.

\subsubsection{Determination of the Geometric Trajectory Constraints}
As illustrated in Fig.~\ref{fig:ULA_airy_Design}, to ensure the Airy beam effectively bends over the obstacle while ensuring the main lobe reaches the receiver, the beam trajectory must be constrained to pass through two critical spatial coordinates: the waypoint $(z_b, x_s)$ above the blockage and the target point $(z_r, x_c)$ on the receiver plane.
First, the waypoint $(z_b, x_s)$ is determined by the physical dimension of the obstacle and a safety margin $d_s$, given by $x_s = x_b + d_s$, where $x_b$ denotes the coordinate of the blockage edge. 
Second, the target point $(z_r, x_c)$ is selected to align the beam propagation direction with the geometry of the LoS tunnel. We assume the transmitter and receiver arrays have physical apertures of $D_t$ and $D_r$, respectively, and are distributed symmetrically around their centers. The slope $k_{upper}$ of the line connecting the upper boundaries of the transmit and receive arrays is calculated as $k_{upper} = \frac{(x_r + D_r/2) - D_t/2}{z_r}$, where $x_r$ is the center coordinate of the receiver array and the transmitter is centered at the origin. The target point $x_c$ is then determined as the intersection of the receiver plane and the line passing through the waypoint $(z_b, x_s)$ with the slope $k_{upper}$, which yields
\begin{equation}
x_c =  x_s + \frac{x_r + 0.5(D_r - D_t)}{z_r} (z_r - z_b).
\end{equation}
Given that the trajectory is formed by the constructive interference of rays emitted from the finite antenna aperture, the maximum propagation distance of an Airy beam is fundamentally constrained by the array boundaries. Consequently, this geometric configuration ensures the beam adheres to its prescribed path, enabling it to successfully circumvent the obstacle while remaining within the physical support limits of the antenna array.

\subsubsection{Closed-Form Derivation}
After obtaining the geometric trajectory constraints, the Airy beam design problem can be formulated as 
\begin{equation} \begin{aligned} \{B_{{opt}}, F_{{opt}}, \theta_{{opt}}\} = & \arg \max_{B, F, \theta} |E(x_c, z_r)| \\ \text{s.t. } & x(z_b) = x_s ,\ x(z_r) = x_c. \end{aligned} \end{equation}
The objective is to find the optimal parameters that force the beam trajectory through the waypoint and target point, enabling the beam to bend over the obstacle and achieve peak field magnitude at the receiver.
With the two anchor points $(z_b, x_s)$ and $(z_r, x_c)$ determined, the trajectory of the Airy beam is constrained by the following boundary conditions:
\begin{equation}
\begin{cases}
x_s = -\xi_{peak}\lambda z_b B - \sin\theta z_b - \frac{(\frac{1}{z_b}-\frac{1}{F})^2 - S_I^2}{16\lambda\pi^2 B^3} z_b, \\
x_c = -\xi_{peak}\lambda z_r B - \sin\theta z_r - \frac{(\frac{1}{z_r}-\frac{1}{F})^2 - S_I^2}{16\lambda\pi^2 B^3} z_r.
\end{cases}
\label{eq:boundary_conditions}
\end{equation}
By eliminating the common steering angle term $\sin\theta$ from \eqref{eq:boundary_conditions}, a deterministic relationship between the focusing distance $F$ and the bending parameter $B$ is established
\begin{equation}
\frac{1}{F} = Q_1 + Q_2 B^3,
\label{eq:F_B_relation}
\end{equation}
where the coefficients are given by $Q_1 = \frac{1}{2}(\frac{1}{z_r} + \frac{1}{z_b})$ and $Q_2 = \frac{8\lambda\pi^2(\frac{x_c}{z_r} - \frac{x_s}{z_b})}{(\frac{1}{z_r} - \frac{1}{z_b})}$. 
Substituting this relationship into the electric field magnitude expression derived in \eqref{eq:electric_amplitude}, the optimization problem is reduced to a one-dimensional search for $B$ that maximizes the received signal strength. To obtain a closed-form solution, we analyze the stationary point of the log-amplitude objective function $J(B) = \ln|E(x_c, z_r)|$. Neglecting the constant terms, the expression can be given by
\begin{equation}
    J(B)=-\ln{B}-\frac{K_2}{B^2}-\frac{K_6}{B^6}.
\end{equation}

By substituting $K_2$, $K_6$ and differentiating with respect to $B$, we set $J^{'}(B)=0$ to find the extreme point. The resulting equation can be expressed as
\begin{equation}
    \begin{aligned}
            B^6&+\frac{2\xi_{peak}}{(2\pi\omega_0)^2}B^4+\frac{6(\frac{\pi}{\lambda})^2GQ_2}{(2\pi)^6\omega_0^2}B^3\\&-\frac{6}{3(2\pi)^6\omega_0^6}-\frac{6(\frac{\pi}{\lambda})^2G^2}{(2\pi)^6\omega_0^2}=0.
    \end{aligned}
   \label{eq:B_derivative_zero} 
\end{equation}
In the regime where the cubic phase dominates, the term of $B^4$ in \eqref{eq:B_derivative_zero} has a negligible impact on the root location compared to the $B^6$ and $B^3$ terms and can be eliminated for simplicity. By defining $T=B^3$, the sixth-order polynomial equation turns into a tractable quadratic form that can be expressed as
\begin{equation} T^2 + P_1 T + P_2 = 0, \end{equation}
where the coefficients are $P_1=\frac{6(\frac{\pi}{\lambda})^2GQ_2}{(2\pi)^6\omega_0^2}$ and $P_2=-\frac{6}{3(2\pi)^6\omega_0^6}-\frac{6(\frac{\pi}{\lambda})^2G^2}{(2\pi)^6\omega_0^2}$. The optimal curvature $B_{opt}$ is then obtained as
\begin{equation} B_{opt} =\sqrt[3]T = \sqrt[3]{\frac{-P_1 +\sigma \sqrt{P_1^2 - 4P_2}}{2}}, \end{equation} 
where $\sigma$ is chosen from $\{+1,-1\}$ to ensure $B_{opt}$ remains consistent with the required bending direction dictated by the blockage position relative to the receiver. After obtaining $B_{opt}$, $F_{opt}$ and $\theta_{opt}$ can be derived based on \eqref{eq:F_B_relation} and \eqref{eq:boundary_conditions}.

The complete closed-form formulations for the optimal Airy beam parameters $\{B_{opt},F_{opt},\theta_{opt}\}$ in ULA systems are presented in \eqref{eq:closed_form_B}, \eqref{eq:closed_form_F} and \eqref{eq:closed_form_theta}, respectively, which are only dependent on $(z_r,x_c)$, $(z_b,x_s)$ and $\omega_0$. This deterministic relationship implies that once the environmental information is acquired, the optimal beam configuration can be directly computed. This stands in contrast to conventional alternative solutions, such as exhaustive grid searches that require high-dimensional scanning of the $\{B,F,\theta\}$ parameter space, or iterative numerical optimizations involving repeated electromagnetic field simulations. Consequently, this closed-form Airy beam design offers minimal computational complexity and high real-time efficiency, providing a robust theoretical framework for Airy beam beamforming design.

% By solving $\frac{\partial J(B)}{\partial B} = 0$ and simplifying the resulting higher-order equation, the optimal curvature parameter $B_{opt}$ is derived as:
% \begin{equation}
% B_{opt} = \sqrt[3]{\frac{-P_1 \pm \sqrt{P_1^2 - 4P_2}}{2}},
% \label{eq:B_opt}
% \end{equation}
% where $P_1$ and $P_2$ are coefficients derived from the amplitude analysis. The sign is chosen to satisfy the direction indicator $\sigma \in \{1, -1\}$ determined by the blockage position.

% Finally, with $B_{opt}$ obtained, the optimal focusing distance $F_{opt}$ and steering angle $\theta_{opt}$ are directly calculated using \eqref{eq:F_B_relation} and \eqref{eq:boundary_conditions}. This set of parameters $\{B_{opt}, F_{opt}, \theta_{opt}\}$ constitutes the closed-form Airy beam design, which can be directly computed from the spatial coordinates of the transceivers and obstacles without iterative optimization.

\begin{figure*}[!t]
\normalsize
% \begin{equation}
%     % --- 公式 B ---
%      B =\frac{1}{\sqrt[3]{2}}\sqrt[3]{
%     \frac{ - 3(\frac{\pi}{\lambda})^2  \cdot 8\lambda\pi^2 (\frac{x_r}{z_r} - \frac{x_b}{z_b}) }{ (2\pi)^6  } 
%     \pm \sqrt{ 
%         \left( \frac{ 3(\frac{\pi}{\lambda})^2 \cdot 8\lambda\pi^2 (\frac{x_r}{z_r} - \frac{x_b}{z_b}) }{ (2\pi)^6 } \right)^2 
%         +4 \left( \frac{2}{(2\pi)^6 \omega_0^2} +\frac{ 3(\frac{\pi}{\lambda})^2 \left( \frac{1}{z_r} - \frac{1}{z_b} \right)^2 }{ (2\pi)^6 } \right) 
%     } 
% }
%  \label{eq:closed_form_B} 
%  \end{equation}
    % --- 间距 ---
       % \begin{equation} B_{\mathrm{opt}} = \sqrt[3]{ -\frac{3 \left( \frac{\pi}{\lambda} \right)^2 G Q_2}{(2\pi)^6 \omega_0^2} + \sigma \sqrt{ \frac{9 \left( \frac{\pi}{\lambda} \right)^4 G^2 Q_2^2}{(2\pi)^{12} \omega_0^4} +\frac{2}{(2\pi)^6 \omega_0^6} + \frac{6 \left( \frac{\pi}{\lambda} \right)^2 G^2}{(2\pi)^6 \omega_0^2} } } \label{eq:B_opt_expanded} \end{equation}   
       % \begin{equation} B_{\mathrm{opt}} = \sqrt[3]{ -\frac{12\left( \frac{\pi}{\lambda} \right)^2 \lambda\pi^2\left(\frac{x_r}{z_r}-\frac{x_b}{z_b}\right)}{(2\pi)^6 \omega_0^2} + \sigma \sqrt{ \frac{144 \left( \frac{\pi}{\lambda} \right)^4 \lambda^2\pi^4\left(\frac{x_r}{z_r}-\frac{x_b}{z_b}\right)^2}{(2\pi)^{12} \omega_0^4} +\frac{2}{(2\pi)^6 \omega_0^6} + \frac{3 \left( \frac{\pi}{\lambda} \right)^2 \left(\frac{1}{z_r}-\frac{1}{z_b}\right)^2}{2(2\pi)^6 \omega_0^2} } } \label{eq:B_opt_expanded} \end{equation}

       \begin{equation} B_{{opt}} = \sqrt[3]{ -\frac{3 \left( \frac{x_c}{z_r} - \frac{x_s}{z_b} \right)}{16 \lambda \pi^2 \omega_0^2} + \sigma \sqrt{ \left[ \frac{3 \left( \frac{x_c}{z_r} - \frac{x_s}{z_b} \right)}{16 \lambda \pi^2 \omega_0^2} \right]^2 + \frac{2}{(2\pi)^6 \omega_0^6} + \frac{3 \left( \frac{1}{z_r} - \frac{1}{z_b} \right)^2}{128 \lambda^2 \pi^4 \omega_0^2} } } \label{eq:closed_form_B}  \end{equation}

  % \begin{equation} B_{\mathrm{opt}} = \sqrt[3]{ \frac{3 \left( \frac{x_r}{z_r} - \frac{x_b}{z_b} \right)}{16 \lambda \pi^2 \omega_0^2} + \sigma \sqrt{ \frac{9 \left( \frac{x_r}{z_r} - \frac{x_b}{z_b} \right)^2}{256 \lambda^2 \pi^4 \omega_0^4} - \frac{3 \left( \frac{1}{z_r} - \frac{1}{z_b} \right)^2}{128 \lambda^2 \pi^4 \omega_0^2} - \frac{1}{32 \pi^6 \omega_0^6} } } \label{eq:B_opt_final_substituted} \end{equation}

\begin{equation}
     % --- 公式 F ---
     F_{opt} = \frac{1}{ \frac{1}{2}\left(\frac{1}{z_r}+\frac{1}{z_b}\right) + \frac{ 8\lambda\pi^2 (\frac{x_c}{z_r} - \frac{x_s}{z_b}) }{ (\frac{1}{z_r} - \frac{1}{z_b}) } B^3 } \label{eq:closed_form_F} 
    % --- 间距 ---
\end{equation}

   \begin{equation}
        % --- 公式 Theta ---
     \theta_{opt} = \arcsin\left( -\xi_{peak}\lambda B - \frac{x_s}{z_b} - \frac{ \left( \frac{1}{z_b} - \frac{1}{F} \right)^2 - \left(\frac{\lambda}{\pi\omega_0^2}\right)^2 }{ 16\lambda\pi^2 B^3 } \right)  \label{eq:closed_form_theta}
   \end{equation}

\hrulefill
\vspace*{4pt}
\end{figure*}

\section{Closed-Form Airy Beam Design for Uniform Planar Arrays}
\label{sec:UPA}
Most prior works studying Airy beams have been confined to ULA configurations \cite{Chen-2024-Curving,Chen-2025-physics,Yazdnian-2025-NirvaWave,Zhao-2025-WDC,Weng-2025-Learning}. In this section, we expand the communication scenario to the more general UPA architecture deployed on the $xy$-plane with $N_t=N_x\times N_y$ antennas. We provide a comprehensive analysis of the electric field, 3D beam trajectory, and field amplitude, establishing a theoretical connection to the ULA situations. Based on these derivations, we investigate the Airy beam design in UPA systems and propose closed-form phase profile solutions with two Airy beam operation modes.
%把已知收发端以及障碍物的位置，确定airy beam的情况讲一下
\subsection{Electric Field}
To generate a 2D Airy beam, we impose a separable cubic phase modulation on the UPA. The total phase distribution $\phi(x_0, y_0)$ at the transmitter aperture is designed as the superposition of the phase profiles along the two transverse dimensions, which can be expressed as
\begin{equation}
\label{eq:UPA_phase}
\phi(x_0, y_0) = \phi_x(x_0) + \phi_y(y_0),
\end{equation}
where $\phi_x(x_0)$ and $\phi_y(y_0)$ follow the 1D Airy phase definition given in (6), but are configured with independent parameters $(B_x, F_x, \theta_x)$ and $(B_y, F_y, \theta_y)$, respectively.
Similar to the ULA case, the abrupt truncation of a rectangular aperture introduces sharp spectral discontinuities that complicate analytical derivation. To ensure analytical tractability, we employ a 2D Gaussian window $W(x_0, y_0)$ to model the finite aperture of the UPA, which can be expressed as $W(x_0, y_0) = e^{-\frac{x_0^2}{\omega_{0x}^2}} \cdot e^{-\frac{y_0^2}{\omega_{0y}^2}}$,
% \begin{equation}
% W(x_0, y_0) = e^{-\frac{x_0^2}{\omega_{0x}^2}} \cdot e^{-\frac{y_0^2}{\omega_{0y}^2}},
% \end{equation}
where $\omega_{0x}$ and $\omega_{0y}$ related to the physical aperture along the $x$ and $y$ axes. Consequently, the initial electric field at the transmitter plane $z=0$ is given by $\widetilde{E}(x_0, y_0, 0) = W(x_0, y_0)e^{j\phi(x_0, y_0)}$.
Under the Fresnel approximation, the propagation of the electric field in 3D space is governed by the diffraction integral
\begin{equation}
E(x, y, z) = \frac{e^{jkz}}{j\lambda z} \iint_{-\infty}^{\infty} \widetilde{E} e^{j\frac{k}{2z}[(x-x_0)^2 + (y-y_0)^2]} \, dx_0 \, dy_0.
\end{equation}
Substituting the initial field expression into the integral, we observe that the exponential kernel $e^{j\frac{k}{2z}[(x-x_0)^2 + (y-y_0)^2]}$ can be factored into $x$-dependent and $y$-dependent terms. This allows the double integral to be decoupled into the product of two independent 1D integrals
\begin{align}
E(x, y, z) &= \frac{e^{jkz}}{j\lambda z} \left[ \int_{-\infty}^{\infty}  e^{j\phi_x(x_0)} e^{j\frac{k}{2z}(x-x_0)^2}e^{-\frac{x_0^2}{\omega_{0x}^2}} \, dx_0 \right] \nonumber \\
&\quad \times \left[ \int_{-\infty}^{\infty} e^{j\phi_y(y_0)} e^{j\frac{k}{2z}(y-y_0)^2}e^{-\frac{y_0^2}{\omega_{0y}^2}} \, dy_0 \right].
\end{align}
Remarkably, each bracketed term corresponds exactly to the 1D propagation integral derived for the ULA case in (9). By invoking~\eqref{eq:electric_field}, the final closed-form expression for the electric field of a UPA Airy beam can be written as the product of two orthogonal 1D field components:
\begin{equation}
\label{eq:UPA_E_field}
E_{UPA}(x, y, z) = \frac{1}{j \lambda z} \cdot \Psi_x(x, z) \cdot \Psi_y(y, z) \cdot e^{jkz},
\end{equation}
where $\Psi_{u}(u, z)$ (for $u \in \{x, y\}$) represents the kernel of the 1D solution derived in Sec.~\ref{sec:ULA}, which can be given by
\begin{equation}
\Psi_{u}(u, z) = \frac{1}{B_u} e^{j\frac{\pi}{\lambda z}u^2} e^{j\left(\frac{2C_{2,u}^3}{3A_u^2} - \frac{C_{1,u}C_{2,u}}{A_u}\right)} \text{Ai}(\xi_u),
\end{equation}
where the parameters $A_u$, $C_{1,u}$, $C_{2,u}$, and the Airy function argument $\xi_u$ are defined identically to the ULA case in (13) and (17), utilizing the respective dimensional parameters. 

\subsection{Trajectory}
Based on the derived electric field expression in \eqref{eq:UPA_E_field}, the intensity distribution of the Airy beam in a UPA system is proportional to the product of the squared Airy functions in the two transverse dimensions, which is expressed as
\begin{equation}
I(x, y, z) \propto |E_{UPA}(x, y, z)|^2 \propto \left| \text{Ai}(\xi_x) \right|^2 \cdot \left| \text{Ai}(\xi_y) \right|^2.
\end{equation}
Since the Airy function $\text{Ai}(\cdot)$ achieves its global maximum at $\xi_{peak} \approx -1.0188$, the peak intensity of the main lobe in 3D space is located at the spatial coordinate where both $\xi_x$ and $\xi_y$ simultaneously satisfy the peak condition.
Consequently, the 3D trajectory can be mathematically decoupled into two independent parametric equations relative to the propagation distance $z$. By solving $\mathcal{R}\{\xi_x\} = \xi_{peak}$ and $\mathcal{R}\{\xi_y\} = \xi_{peak}$ using the explicit form of $\xi$ derived in~\eqref{eq:electric_field}, we obtain the closed-form trajectory equations for the UPA Airy beam, which can be expressed as 
\begin{equation}
\label{eq:3D_Trajectory}
\begin{cases}
x(z) = -\xi_{peak}\lambda z B_x - \sin\theta_x z - \frac{S_{R,x}^2 - S_{I,x}^2}{16\lambda\pi^2 B_x^3}z, \\
y(z) = -\xi_{peak}\lambda z B_y - \sin\theta_y z - \frac{S_{R,y}^2 - S_{I,y}^2}{16\lambda\pi^2 B_y^3}z,
\end{cases}
\end{equation}
where $S_{R,u} = \frac{1}{z} - \frac{1}{F_u}$ and $S_{I,u} = \frac{\lambda}{\pi \omega_{0u}^2}$ for $u \in \{x, y\}$.
This result reveals a fundamental property of Airy beamforming in UPA systems, i.e., trajectory decoupling. The bending behavior in the azimuth plane governed by $x(z)$ is solely determined by the parameters $\{B_x, F_x, \theta_x\}$, while the elevation bending governed by $y(z)$ is independently controlled by $\{B_y, F_y, \theta_y\}$. This implies that the projection of the 3D trajectory onto the $x$-$z$ or $y$-$z$ plane is identical to the trajectory of a 1D Airy beam configured with the corresponding parameters.

% To validate the proposed solution, the simulated 3D propagation trajectory is illustrated in Fig.~\ref{fig:UPA_trajectory}. It is observed that the red solid curve representing the theoretical 3D trajectory precisely intersects the high-energy main lobe of the simulated electric field at $z=1.2$ m plane, verifying the spatial accuracy. Furthermore, the projections of the 3D trajectory onto the $x$-$z$ and $y$-$z$ planes are identical to the independent 1D profiles, confirming that the 3D beamforming can be effectively decoupled into two orthogonal dimensions.

% \begin{figure}[t]
%     \centering
%     \includegraphics[width=0.8\linewidth]{closed-form_Airy/images/UPA/UPA_trajectory.png} % 使用 \linewidth 自适应栏宽
%     \caption{Airy Beam trajectory in UPA system with $D_x=D_y=0.2721$, $d=\lambda/2, B_x=B_y=5$, $F_x=1$, $F_y=0.5$, $\theta_x=-0.0262$, $\theta_y=0.0175$.}
%        \label{fig:UPA_trajectory}
%     \end{figure}
\subsection{Electric Field Magnitude}
Based on the closed-form electric field derived in \eqref{eq:UPA_E_field}, the magnitude of the UPA Airy beam along its main trajectory can be evaluated as
\begin{equation}
|E_{UPA}(z)| = \left| \frac{1}{\lambda z} \Psi_x(z) \Psi_y(z) \right|.
\end{equation}
By substituting the 1D magnitude approximation derived in \eqref{eq:electric_amplitude} into the separable components $\Psi_x$ and $\Psi_y$, the normalized electric field magnitude is given by
\begin{equation} \begin{aligned} |E_{UPA}|  & \approx\frac{C_{Ai}^2}{\lambda z B_x B_y} \exp \Biggl[ -\left( \frac{K_{2,x}}{B_x^2} + \frac{K_{6,x}}{B_x^6} \right) \\ & -\left( \frac{K_{2,y}}{B_y^2} + \frac{K_{6,y}}{B_y^6} \right) \Biggr], \end{aligned} \label{eq:UPA_mag_formula} \end{equation}
where $K_{2,u}$ and $K_{6,u}$ ($u \in \{x, y\}$) are the coefficients defined in \eqref{eq:Im_phi_c}, calculated using the specific parameters of each dimension.
\subsection{Airy Beam Design in UPA}
The primary objective of the Airy beam design is to determine the optimal configuration that maximizes the received signal strength at the target location while satisfying the blockage avoidance requirements.

% For a UPA system, we define the log-amplitude objective function $J_{UPA} = \ln|E_{UPA}|$ based on the magnitude derived in \eqref{eq:UPA_mag_formula}. Neglecting the constant terms, the optimization problem can be formulated as
% \begin{align}
% \max_{B_x, B_y} J_{UPA} &= \max_{B_x, B_y} \left[ \left( -\ln B_x - \frac{K_{2,x}}{B_x^2} - \frac{K_{6,x}}{B_x^6} \right) \right. \nonumber \\
% &\quad \left. + \left( -\ln B_y - \frac{K_{2,y}}{B_y^2} - \frac{K_{6,y}}{B_y^6} \right) \right] \nonumber \\
% &= \max_{B_x} J(B_x) + \max_{B_y} J(B_y).
% \end{align}
In the ULA scenario, the trajectory design is confined to a 2D plane, rendering the determination of the waypoint $(z_b,x_s)$ straightforward and unique. In contrast, the 3D configuration of a UPA introduces an additional degree of freedom, making the selection of the spatial waypoint $(z_b,x_s,y_s)$ a critical design variable. Consequently, the design of the Airy beam in a UPA system should be decoupled into two sequential tasks. First, identifying the optimal bending dimension and the corresponding waypoint to minimize path deviation. Second, synthesizing the Airy beam phase profile based on the determined waypoint.
\begin{figure}[t]
    \centering
    \subfigure[\text{x} dimension.]{\includegraphics[width=0.8\linewidth]{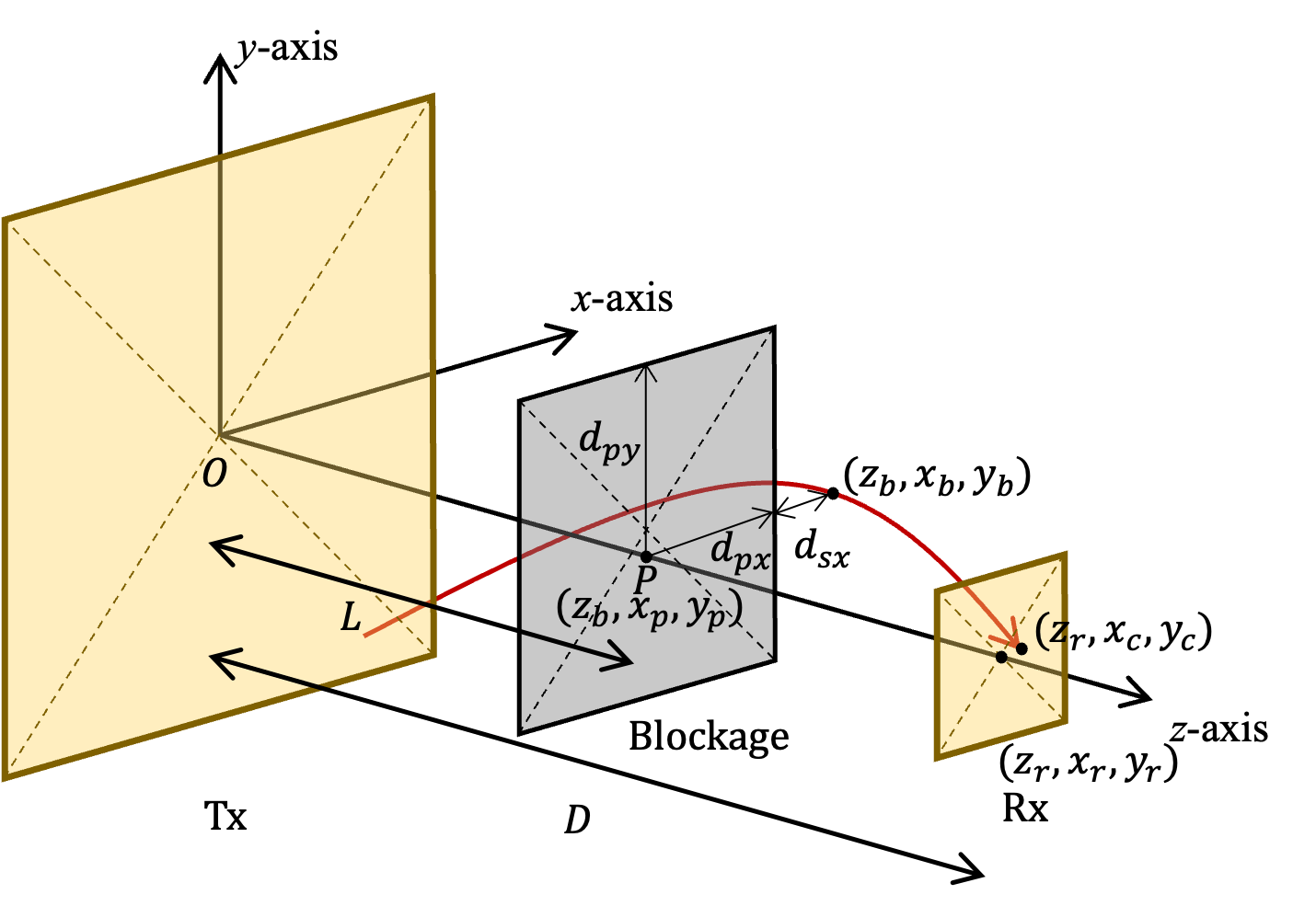}   \label{fig:UPA_x}}
     \subfigure[\text{y} dimension.]{\includegraphics[width=0.8\linewidth]{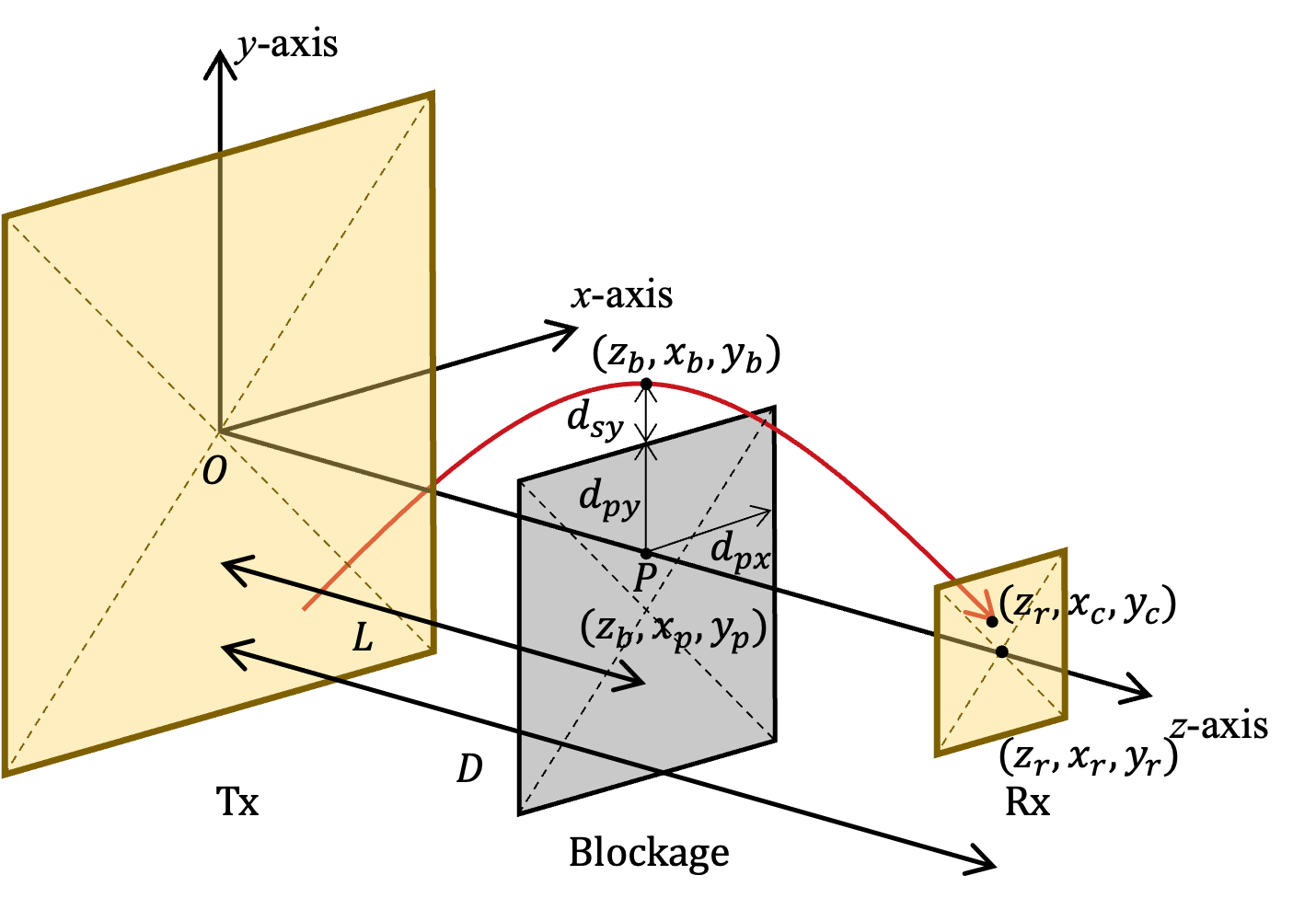}   \label{fig:UPA_y}}
   \caption{Airy beam design in UPA communication systems.}
   \label{fig:UPA_design}
\end{figure} 

To determine the optimal waypoint, we analyze the geometric relationship between the LoS path and the blockage. As illustrated in Fig.~\ref{fig:UPA_design}, although the obstacle blocks the LoS link in both transverse dimensions, it is generally inefficient to induce beam curvature on both dimensions. Instead, it is sufficient to bend the beam primarily along the dimension that requires the minimum spatial deviation to bypass the blockage. Specifically, let the z-axis represent the LoS path connecting the centers of the Tx and Rx, which intersects the blockage plane at point P with $(z_b,x_p,y_p)$. We calculate the required deviation distances from point P to the nearest obstacle boundaries along the x and y axes, denoted as $d_{px}$ and $d_{py}$, respectively. 
The optimal bending dimension is selected by comparing these deviations. The dimension with the smaller deviation is chosen. For instance, in Fig.~\ref{fig:UPA_x}, since $d_{px}<d_{py}$, the x-axis is selected as the bending dimension. Incorporating a safety margin $d_{sx}$, the coordinates of the waypoint $(z_b,x_s,y_s)$ are determined, where the coordinate in the bending dimension is set as $x_s=x_p+d_{px}+d_{sx}$, while the coordinate in the non-bending dimension remains unchanged, i.e., $y_s=y_p$. Conversely, Fig.~\ref{fig:UPA_y} illustrates the scenario where the y-dimension is selected, resulting in a waypoint determined by $y_s=y_p+d_{py}+d_{sy}$.

Once the waypoint is established, the UPA Airy beam parameters can be synthesized. Given the decomposable nature of the 2D Airy beam, two beamforming modes are proposed.
\subsubsection{Hybrid Focusing-Airy Mode} In this mode, the Airy phase profile is applied solely to the selected bending dimension to navigate around the obstacle, while a standard focusing phase is employed in the orthogonal dimension to concentrate energy. Assuming the x-axis is the bending dimension, the design problem is reduced to a 1D Airy beam synthesis for the x-axis parameters $\{B_x,F_x,\theta_x\}$, calculated using \eqref{eq:closed_form_B}-\eqref{eq:closed_form_theta} based on the waypoint $(z_b,x_s)$ and received point $(z_r,x_c)$. Simultaneously, the y-dimension is configured with a focusing phase targeting the received projection based on $(z_r,y_r)$.
\subsubsection{Dual Airy Mode} Alternatively, Airy phase profiles can be applied to both dimensions. This mode provides flexibility for complex obstacle geometries but requires independent parameter calculations. Specifically, the parameters for the x-dimension $\{B_x,F_x,\theta_x\}$ and the y-dimension $\{B_y,F_y,\theta_y\}$ are calculated using \eqref{eq:closed_form_B}-\eqref{eq:closed_form_theta} based on the points $(z_b,x_s)$, $(z_r,x_c)$ and $(z_b,y_s)$, $(z_r,y_c)$, respectively.

% \begin{figure*}[!t] % [!t] 尝试将公式放在页首
% \normalsize % 确保字体大小正常

% \begin{equation}
% B=\Bigg(\frac{-\frac{6(\frac{\pi}{\lambda})^2(\frac{1}{z_r}-\frac{1}{2}(\frac{1}{z_r}+\frac{1}{z_b}))8\lambda\pi^2(\frac{x_r}{z_r}-\frac{x_b}{z_b})}{(2\pi)^6(\frac{1}{z_r}-\frac{1}{z_b})}\pm\sqrt{(\frac{6(\frac{\pi}{\lambda})^2(\frac{1}{z_r}-\frac{1}{2}(\frac{1}{z_r}+\frac{1}{z_b}))8\lambda\pi^2(\frac{x_r}{z_r}-\frac{x_b}{z_b})}{(2\pi)^6(\frac{1}{z_r}-\frac{1}{z_b})})^2-4(-\frac{6}{3(2\pi)^6\omega_0^2}-\frac{6(\frac{\pi}{\lambda})^2(\frac{1}{z_r}-\frac{1}{2}(\frac{1}{z_r}+\frac{1}{z_b}))^2}{(2\pi)^6})}}{2}\Bigg)^3
% \end{equation}

% \begin{equation}
% {F}=\frac{1}{\frac{1}{2}(\frac{1}{z_r}+\frac{1}{z_b})+\frac{8\lambda\pi^2(\frac{x_r}{z_r}-\frac{x_b}{z_b})}{(\frac{1}{z_r}-\frac{1}{z_b})} B^3}
% \end{equation}

% \begin{equation}
%     \theta = \arcsin\bigg(-\xi_{peak}\lambda B - \frac{x_b}{z_b} - \frac{\left( \frac{1}{z_b} - \frac{1}{F} \right)^2 - S_I^2}{16\lambda\pi^2B^3}\bigg) 
% \end{equation}

% \hrulefill % 加上一条分割线，这是 IEEE 的惯例
% \vspace*{4pt} % 调整分割线和正文的间距
% \end{figure*}

%给出非均匀稀疏阵列下旁瓣消失的结果图

\section{SIMULATION RESULTS}
\label{sec:simulation}
%在simulation里面给出随着blockage上升的时候，beam的变化图，blockage上升的时候，按照我们自己的airy beam选择方案，画出来的beam图

In this section, simulations are provided to evaluate the performance of the proposed closed-form Airy beam design framework. We consider a THz hybrid beamforming system where the Tx and Rx are equipped with either ULA or UPA antenna array. The THz system operates at the carrier frequency of $f_c=140$~GHz with the bandwidth of 1~GHz~\cite{Han-2015-Multi-Ray}.

% \begin{figure*}[t]
%     \centering
%     \subfigure[$N_t=256,N_r=64$.]{\includegraphics[width=0.45\linewidth]{closed-form_Airy/images/simulation/ULA/ULA_SE_height_MISO.png} \label{fig:UPA_x}}
%      \subfigure[$N_t=256,N_r=256$.]{\includegraphics[width=0.45\linewidth]{closed-form_Airy/images/simulation/ULA/ULA_SE_height.png}  \label{fig:UPA_y}}
%    \caption{Spectral efficiency versus height of blockage with different number of received antennas. $D=3$~m, $L=1.5$~m.}
%    \label{fig:UPA_design}
% \end{figure*}

    \begin{figure*}[t]
    \centering
        \subfigure[Quasi-LoS full digital precoder.]{\includegraphics[width=0.32\linewidth]{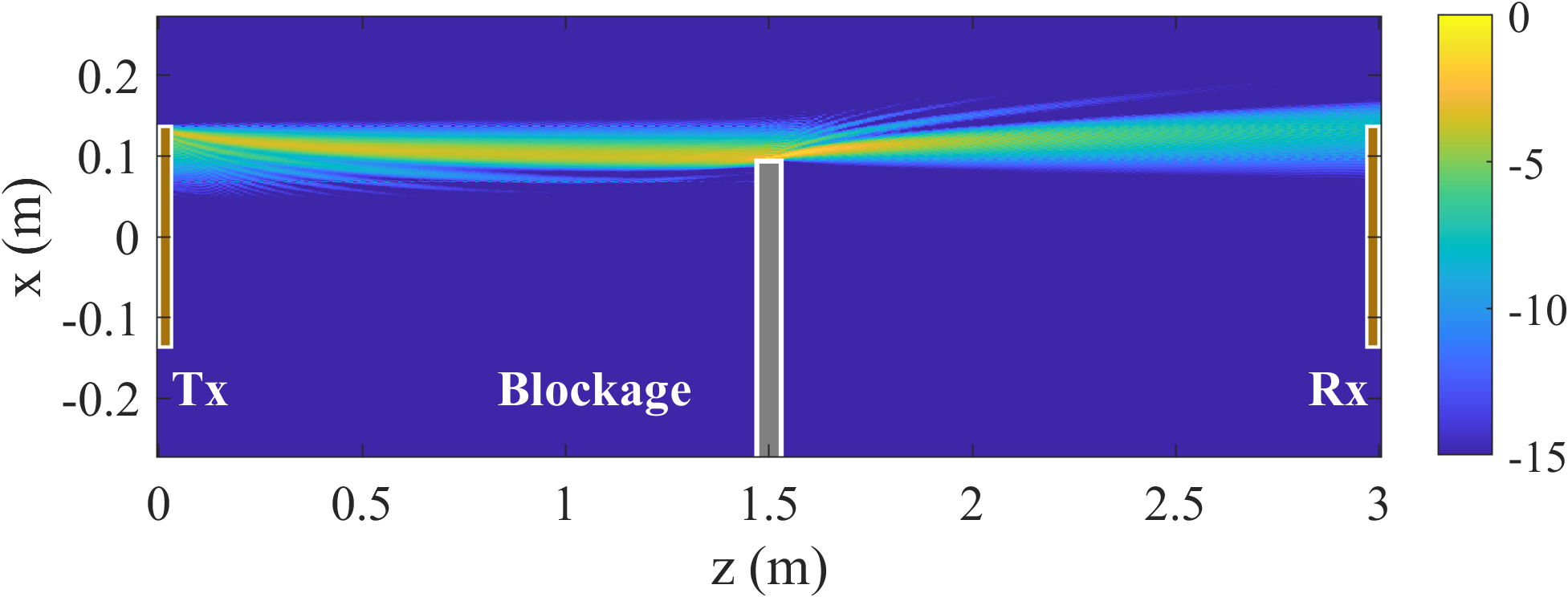} \label{fig:ULA_Blocked precoder}}
       \subfigure[LoS full digital precoder.]{\includegraphics[width=0.32\linewidth]{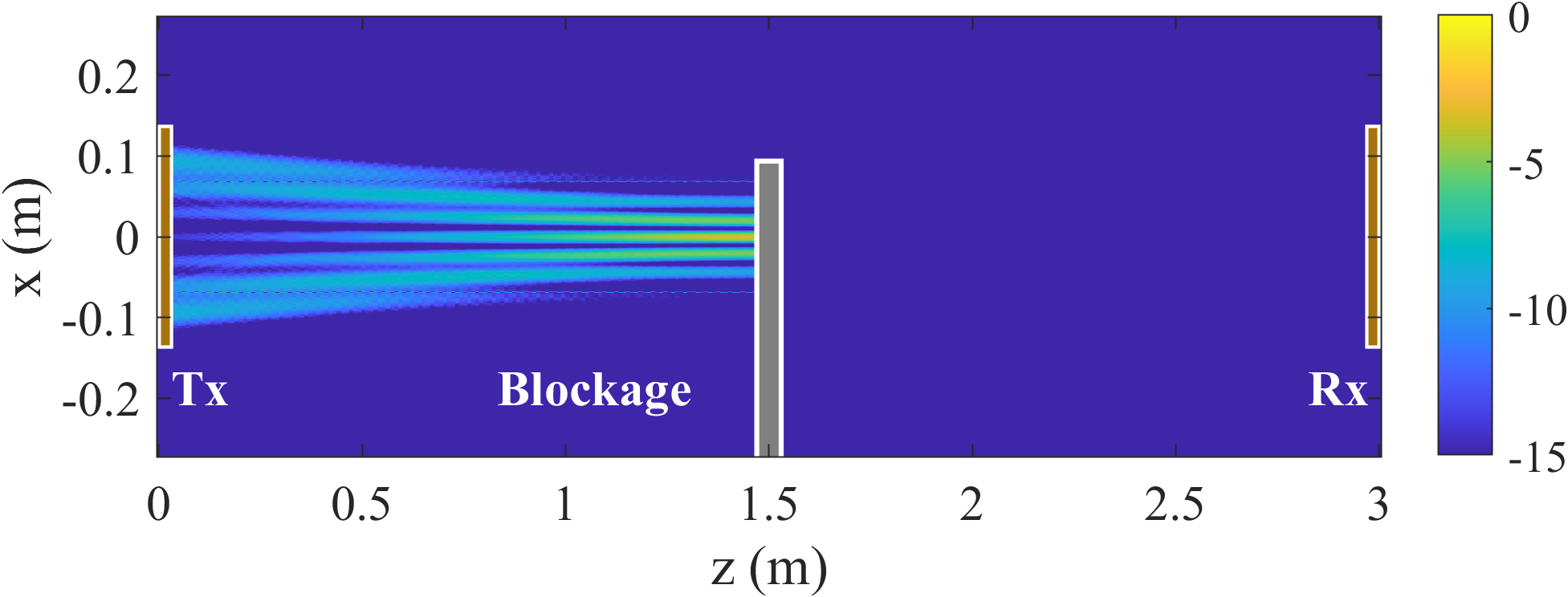} \label{fig:ULA_Non-blocked precoder}} 
 \subfigure[Near-field focusing beam.]{\includegraphics[width=0.32\linewidth]{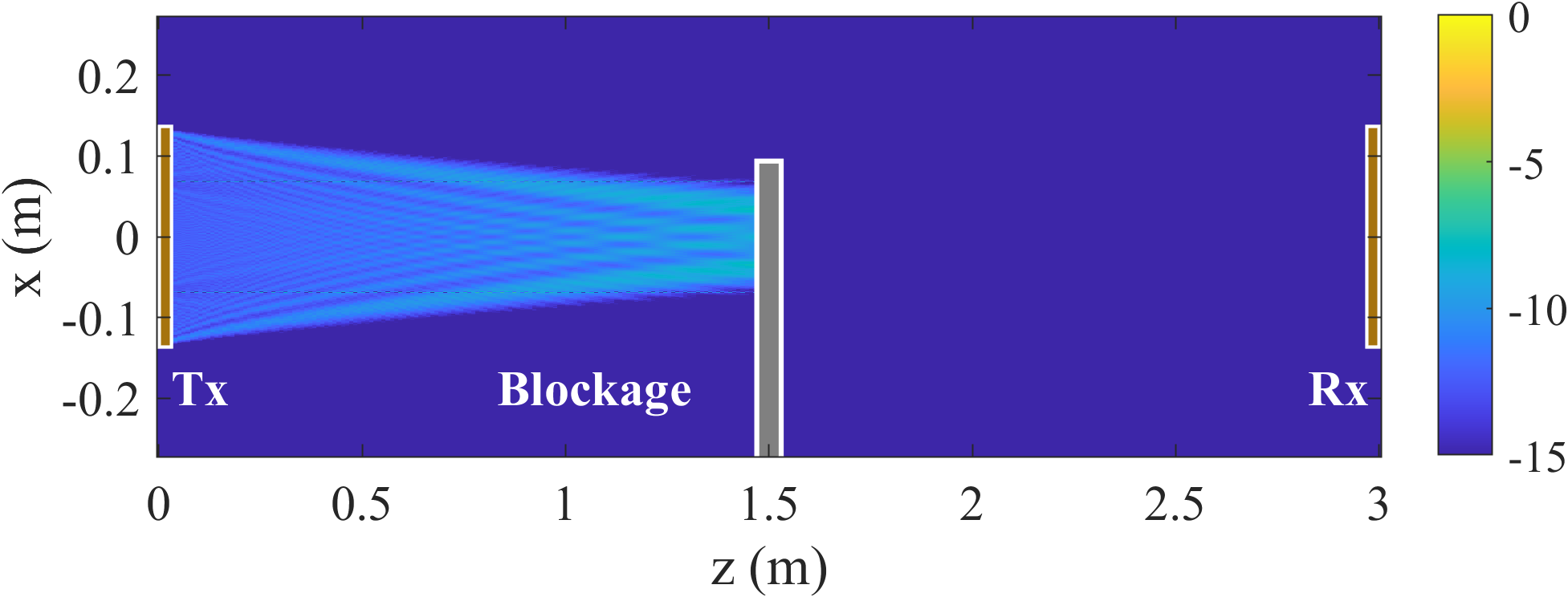} \label{fig:ULA_Focusing beam}}
 \subfigure[Far-field steering beam.]{\includegraphics[width=0.32\linewidth]{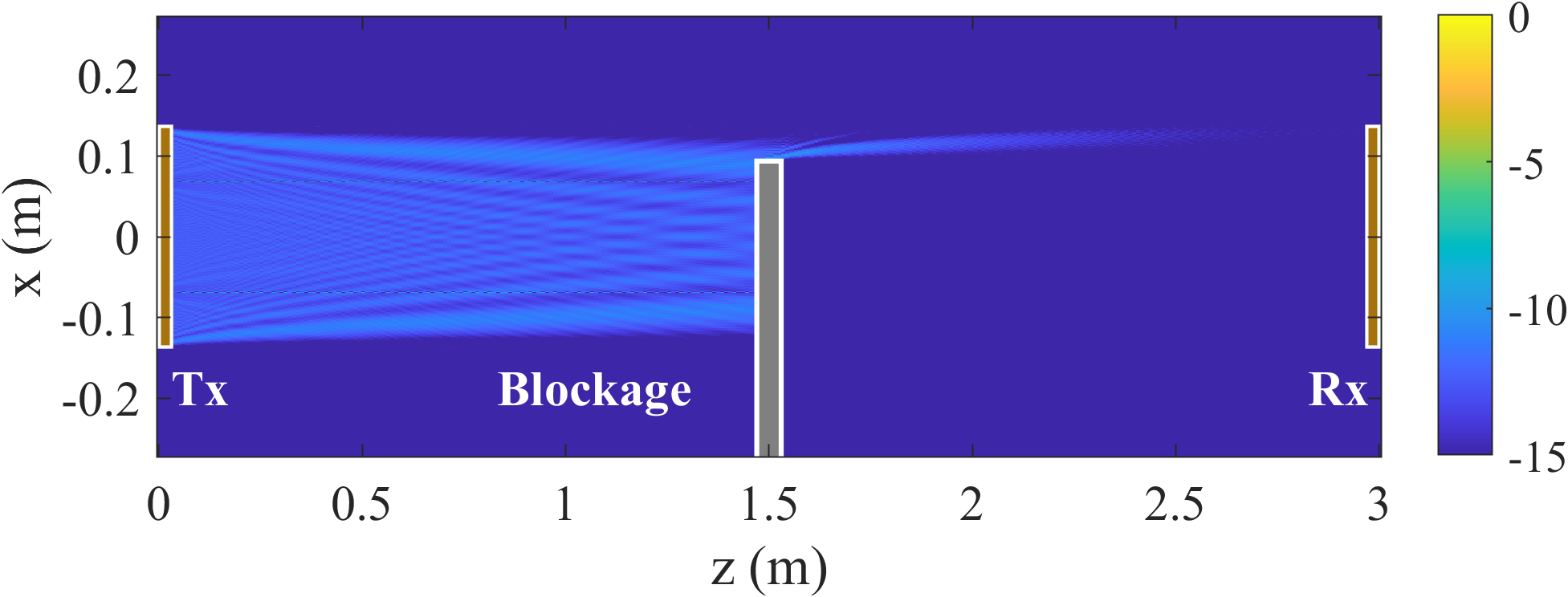} \label{fig:ULA_Steering beam}}
  \subfigure[Exhaustive Airy beam search.]{\includegraphics[width=0.32\linewidth]{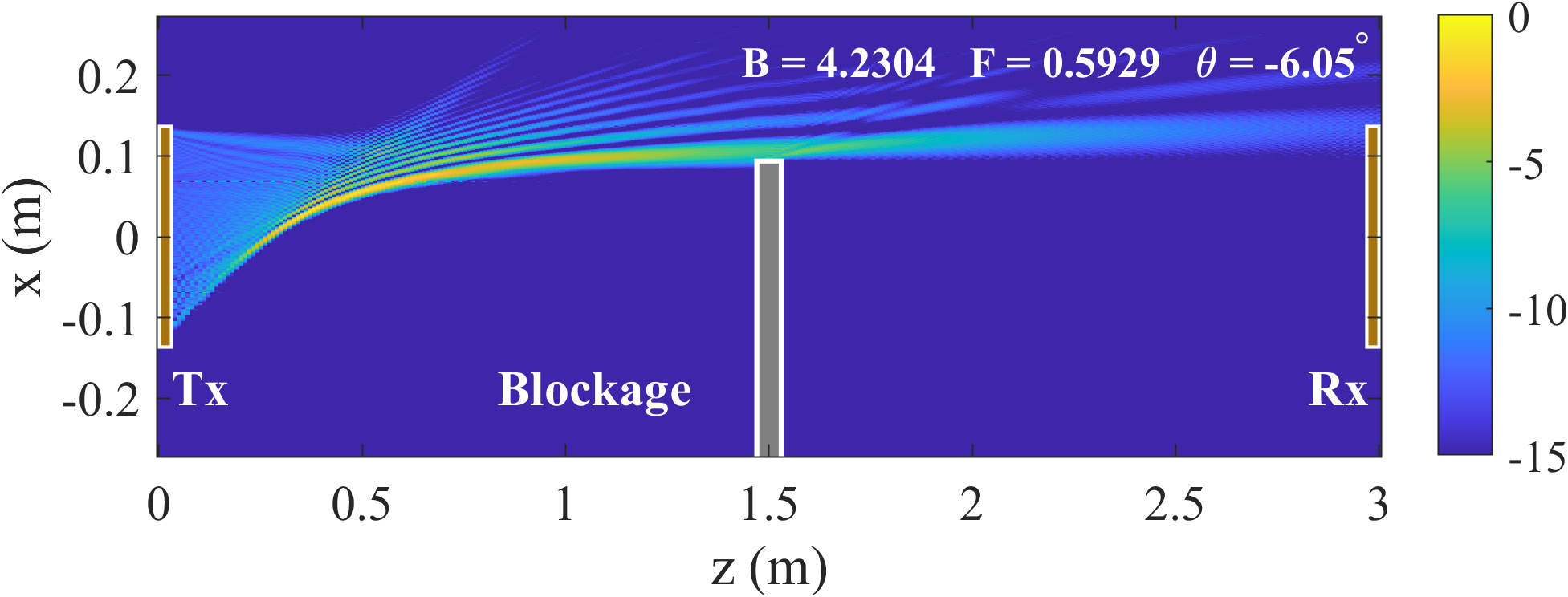} \label{fig:ULA_Exhaustive Airy beam search}}
 \subfigure[Closed-form Airy beam design.]{\includegraphics[width=0.32\linewidth]{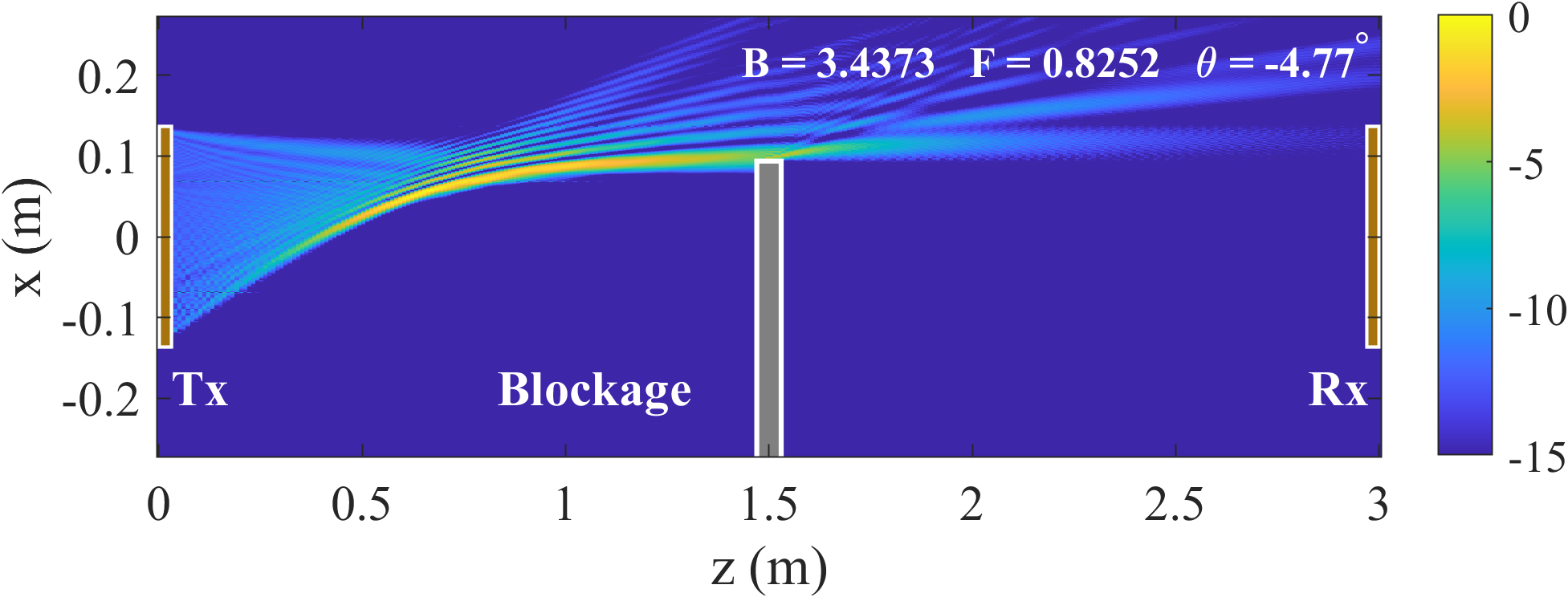} \label{fig:ULA_Closed-form Airy beam design}}

   \caption{Beam comparison with different beamforming schemes. $D=3$~m, $L=1.5$~m and the height of the blockage is 0.071~m ($R_{bl}^{ULA}=75.9\%$).}
   \label{fig:ULA_beam_compare}
\end{figure*}
\begin{figure}[t]
    \centering
    \includegraphics[width=0.8\linewidth]{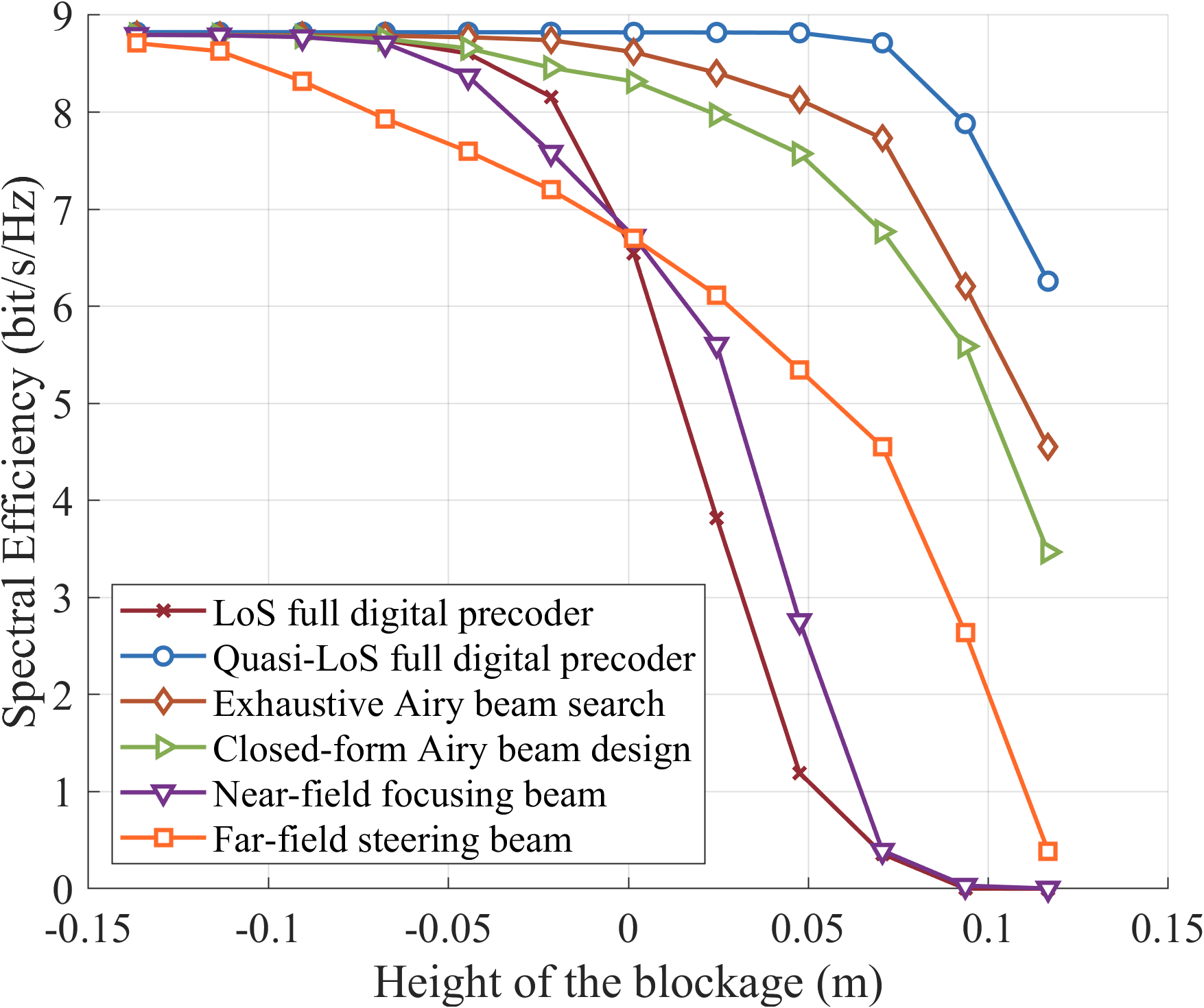} % 使用 \linewidth 自适应栏宽
    \caption{Spectral efficiency versus height of blockage in ULA system. $D=3$~m, $L=1.5$~m, $N_t=N_r=256$, $d=\lambda/2$.}
       \label{fig:ULA_SE}
    \end{figure}

    \begin{figure}[t]
    \centering
    \includegraphics[width=0.8\linewidth]{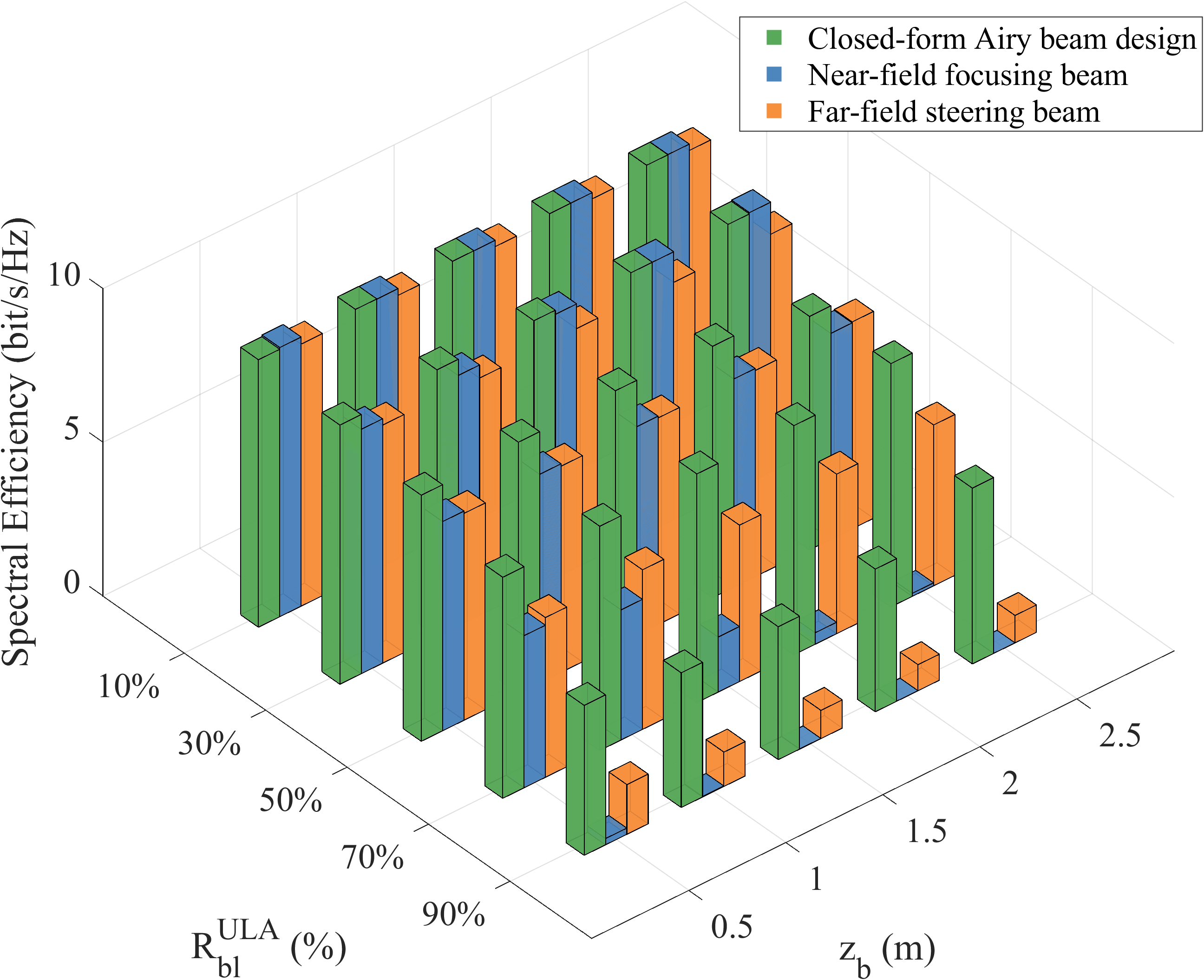} % 使用 \linewidth 自适应栏宽
    \caption{Spectral efficiency versus different position of blockage for different beamforming schemes in ULA system.}
       \label{fig:ULA_2D}
    \end{figure}

\subsection{Airy Beam Design for ULA Systems}

To validate the closed-form of the Airy beam design in ULA systems, we first conduct a simulation where the Tx and Rx are aligned along the z-axis. The center of the Tx antenna array is $(0,0)$ and the center of the Rx antenna array is $(z_r=3,x_r=0)$. $z_b$ is 1.5~m and $x_b$ is variable ranging from the limitation of LoS to NLoS. The Tx and Rx equip ULA with 256 antennas with $d=\lambda/2$. To assess the proposed closed-form Airy beam design scheme, we select the following benchmarks for comparison:

\begin{itemize}
    \item \textbf{LoS full digital precoder}: The full digital precoder is designed using Maximum Ratio Transmission (MRT) and Maximum Ratio Combining (MRC) base on LoS channel, and the spectral efficiency is calculated under the quasi-LoS scenario.

    \item \textbf{Quasi-LoS full digital precoder}: Assuming perfect knowledge of the quasi-LoS channel $\mathbf{H}_b$, the full-digital precoder is designed using MRT and MRC to provide a single-stream preformance upper bound on spectral efficiency in the quasi-LoS scenario.

    \item \textbf{Far-field steering beam} \cite{Alkhateeb-2015-Limited}: This is classical far-field beam steering which is determined by the angle between the Tx and Rx.

    \item \textbf{Near-field focusing beam} \cite{Zhang-2022-Beam-focusing}: This is near-field beam focusing which can be obtained by the position of the Rx.

    \item \textbf{Exhaustive Airy beam search} \cite{Chen-2024-Curving}: This is the upper-bound of using Airy beam which brute force search the optimal combination $\{B,F,\theta\}$ of the Airy beam to maximize the spectral efficiency.
\end{itemize}

We first investigate the propagation behaviors by presenting the electric field intensity distributions for various beamforming schemes in Fig.~\ref{fig:ULA_beam_compare}.
The scenario is configured with a severe blockage height of 0.071~m, corresponding to a blockage ratio of $R_{bl}^{\mathrm{ULA}}=75.9\%$.
In Fig.~\ref{fig:ULA_Blocked precoder}, by performing MRT and MRC on the perfect CSI of the obstructed environment, the beam with full digital precoder serves as the theoretical performance upper bound.
The conventional beamforming schemes, i.e., LoS full digital precoder, near-field focusing beam and far-field steering beam are shown in Fig.~\ref{fig:ULA_Non-blocked precoder}, Fig.~\ref{fig:ULA_Focusing beam} and Fig.~\ref{fig:ULA_Steering beam} which fail to mitigate the influence of blockage. Their main lobes are predominantly intercepted by the obstacle, resulting in minimal energy delivery to the receiver array. In contrast, the Airy beams generated by the proposed closed-form design in Fig.~\ref{fig:ULA_Closed-form Airy beam design} exhibit a curved trajectory which bend over the blockage to deliver a substantial energy to the receiver. Furthermore, the beam pattern calculated by our closed-form solution is nearly identical to that of the exhaustive search in Fig.~\ref{fig:ULA_Exhaustive Airy beam search}, confirming that the derived analytical closed-form expressions can accurately guide the beam trajectory to bypass blockages with negligible computational complexity.

We evaluate the spectral efficiency of different beamforming schemes under varying blockage heights, as illustrated in Fig.~\ref{fig:ULA_SE}. It is evident that the blockage severity plays a critical role in link performance. When the height of blockage is below 0~m, i.e., the blockage ratio under $50\%$, the conventional beamforming schemes, i.e., LoS full digital precoder, near-field focusing beam and far-field steering beam, can still maintain acceptable performance. However, as the blockage ratio exceeds $50\%$, the performance of these traditional schemes deteriorates precipitously due to the severe blockage on the LoS link. In contrast, the Airy beam-based schemes exhibit remarkable robustness, leveraging the self-acceleration property to effectively curve around obstacles. Specifically, in the high-blockage regime ($R_{bl}^{\mathrm{ULA}} > 50\%$), the proposed closed-form Airy beam design demonstrates a significant superiority, achieving an average spectral efficiency gain of approximately 4.63~bit/s/Hz, 4.03~bit/s/Hz and 2.33~bit/s/Hz,  compared to the LoS full digital precoder, near-field focusing beam and far-field steering beam, respectively.
\begin{figure*}[t]
    \centering
        \subfigure[LoS full digital precoder.]{\includegraphics[width=0.3\linewidth]{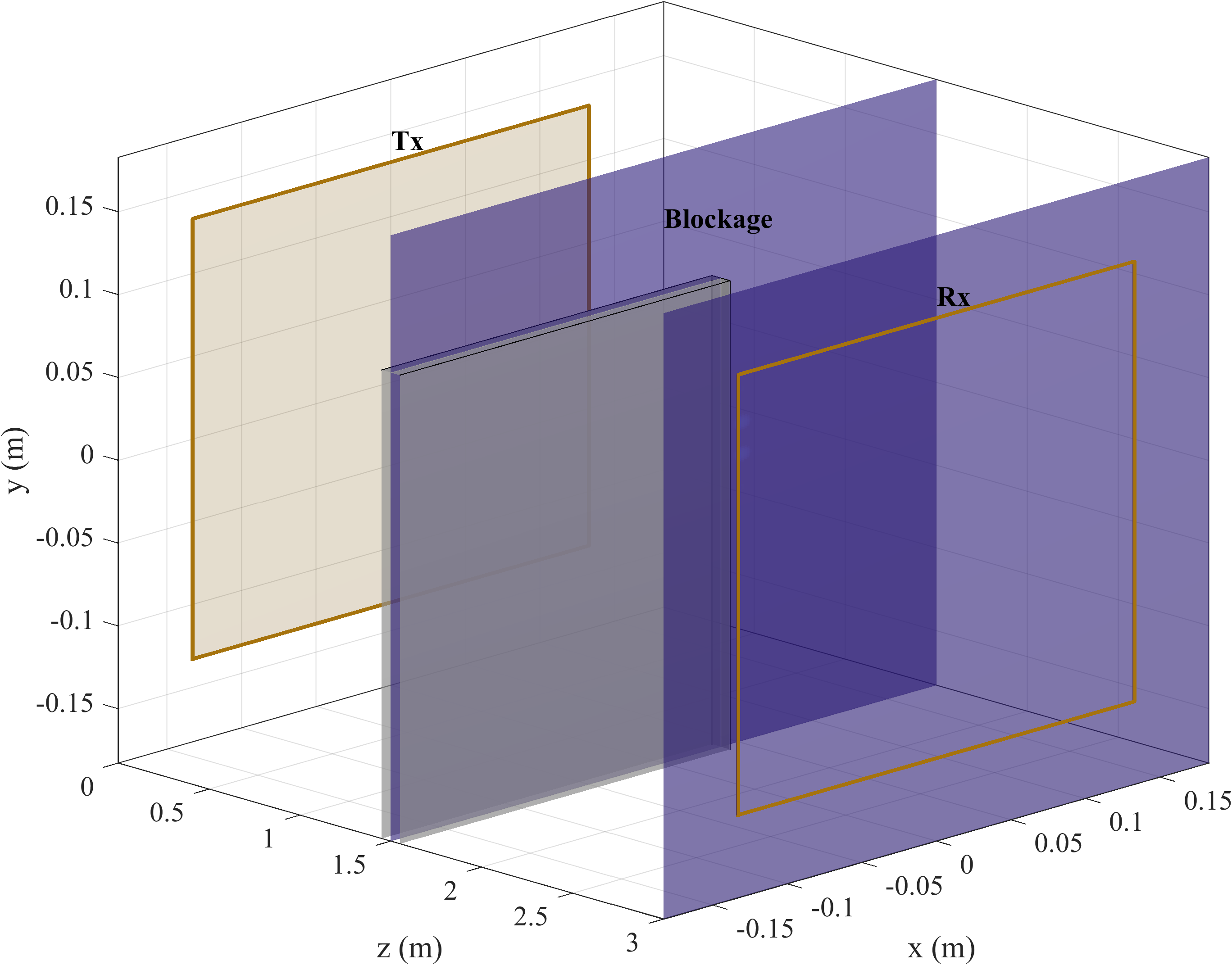} \label{fig:UPA_Non-blocked precoder}}
         \subfigure[Near-field focusing beam.]{\includegraphics[width=0.3\linewidth]{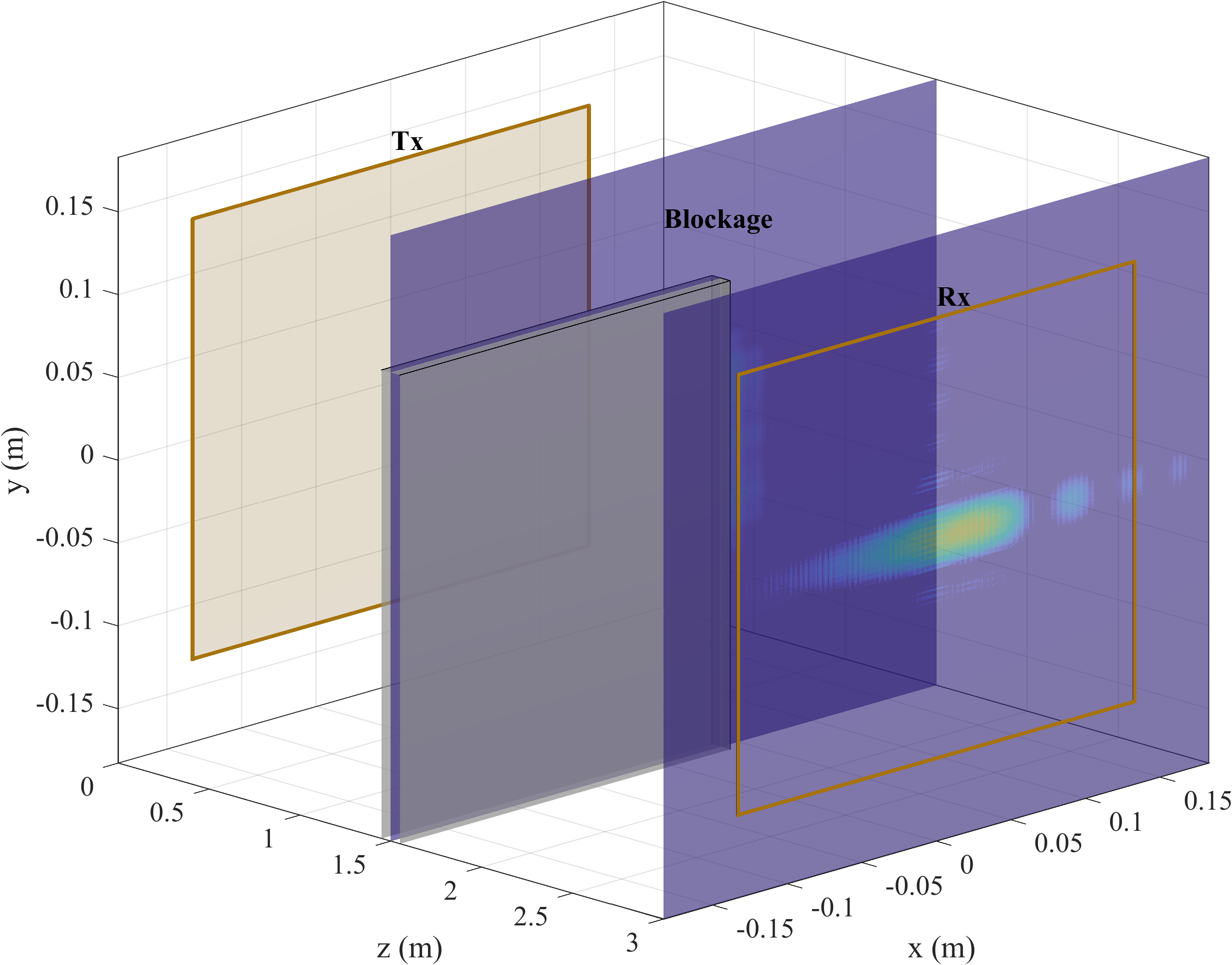} \label{fig:UPA_Focusing beam}}
 \subfigure[Far-field steering beam.]{\includegraphics[width=0.3\linewidth]{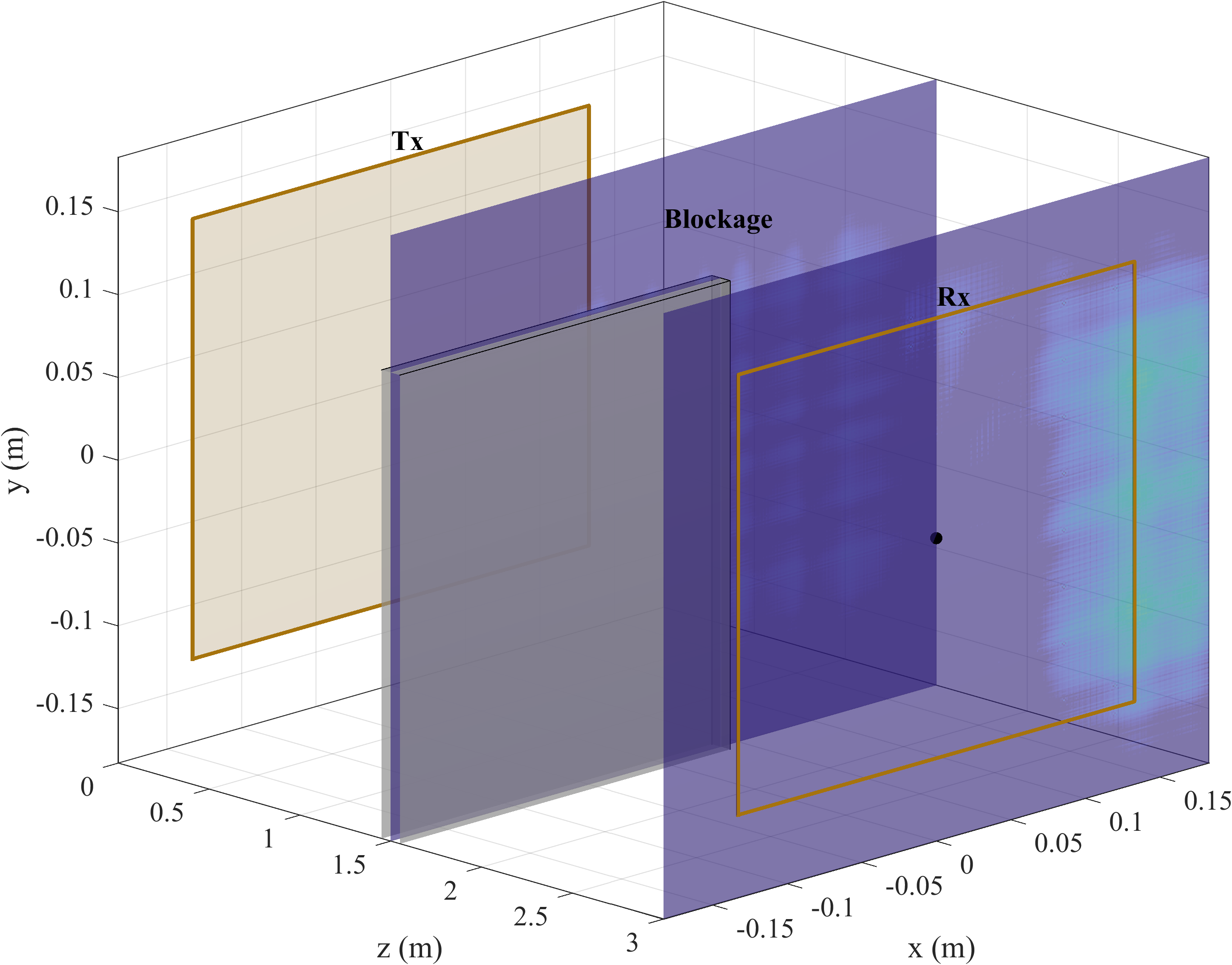} \label{fig:UPA_Steering beam}}
       \subfigure[Exhaustive Airy beam search.]{\includegraphics[width=0.3\linewidth]{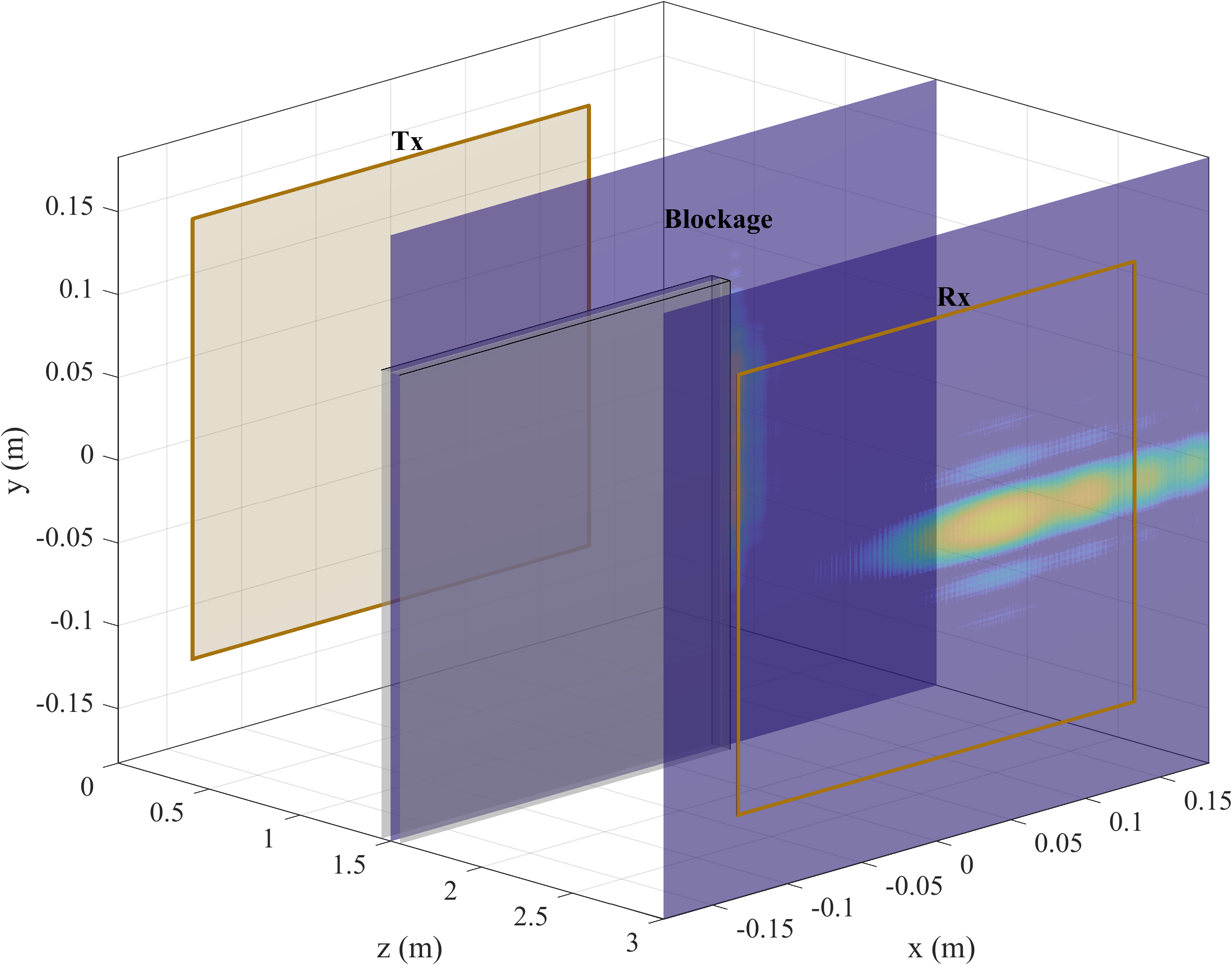} \label{fig:UPA_Exhaustive Airy beam search}} 
        \subfigure[Closed-form Airy beam design (Mode 1).]{\includegraphics[width=0.3\linewidth]{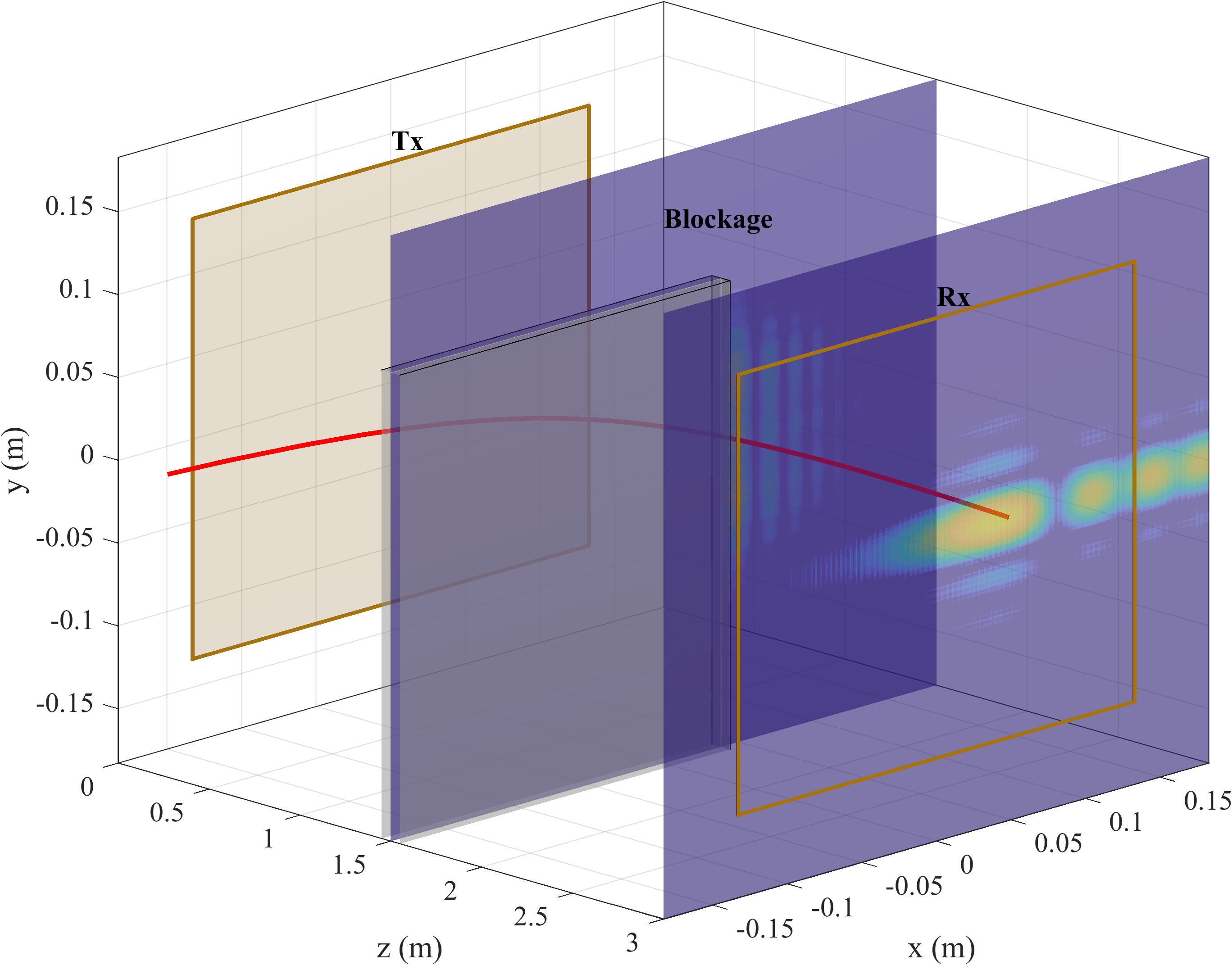} \label{fig:UPA_Closed-form Airy beam design (mode 1)}}
 \subfigure[Closed-form Airy beam design (Mode 2).]{\includegraphics[width=0.3\linewidth]{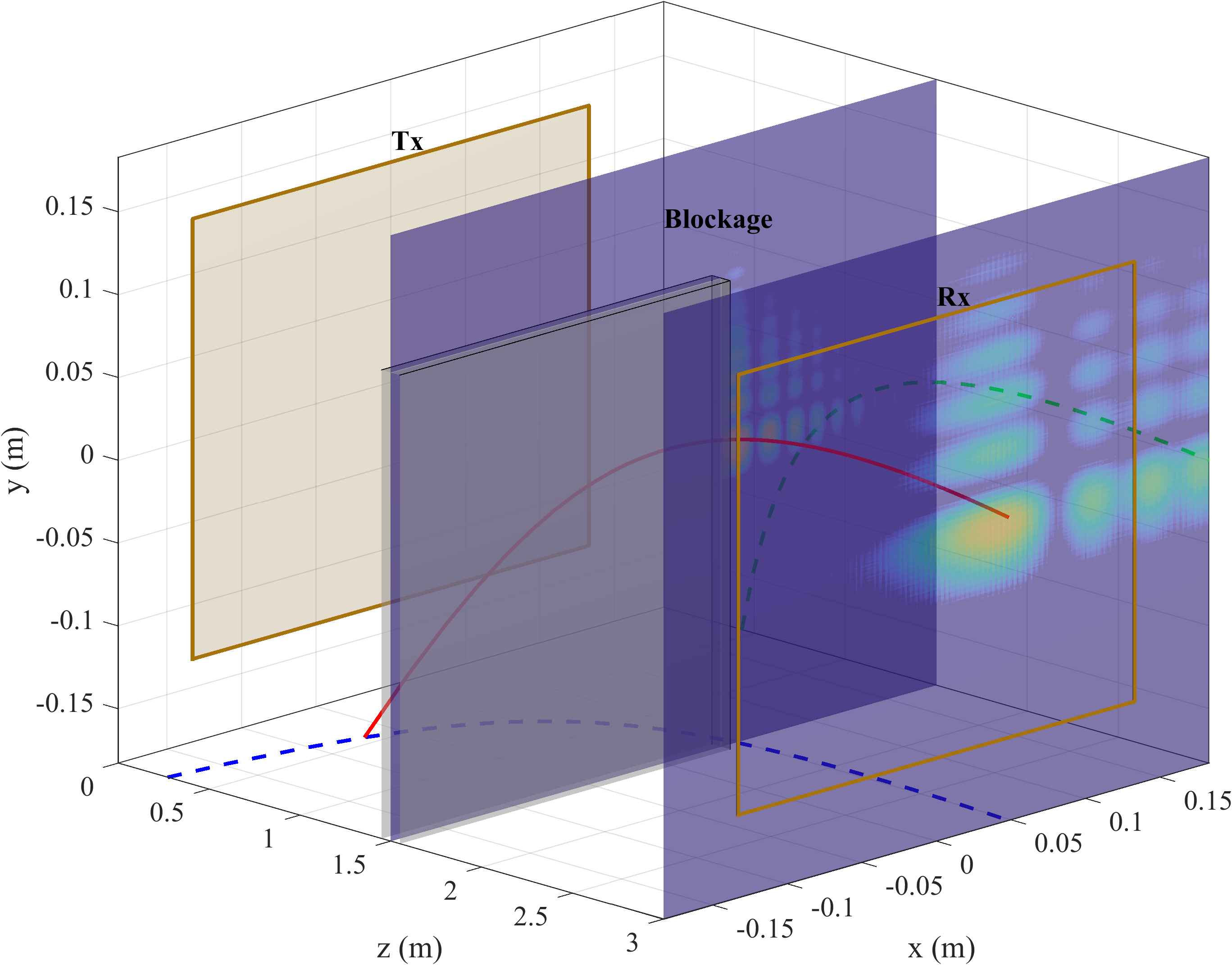} \label{fig:UPA_Closed-form Airy beam design (mode 2)}}

   \caption{Beam comparison with different beamforming schemes. $D=3$~m, $L=1.5$~m and the height of the blockage along x-axis and y-axis is 0.071~m and 0.1~m, respectively ($R_{bl}^{UPA}=65.8\%$).}
   \label{fig:UPA_beam_compare}
\end{figure*}

Notably, a negligible performance gap between the proposed closed-form Airy beam design and the exhaustive Airy beam search is presented in Fig. \ref{fig:ULA_SE}. While the exhaustive search serves as a performance upper bound of the Airy beam, it incurs prohibitive computational complexity and pilot overhead due to the iterative scanning of the parameter space. In contrast, our proposed method directly calculates the Airy beam phase parameters directly based on the position of blockage and receiver, achieving near-optimal spectral efficiency without the latency of beam searching. This validates the performance of our derivation and highlights the efficiency of the closed-form solution for real-time applications.

 To evaluate the generalizability and universality of our proposed closed-form design, Fig.~\ref{fig:ULA_2D} presents a 3D comparison of spectral efficiency across various obstacle distances $z_b$ and blockage ratios $R_{bl}^{\mathrm{ULA}}$. It is evident that the Airy beam demonstrates remarkable resilience across all evaluated scenarios. While conventional focusing and steering beams suffer from serious performance degradation as $R_{bl}^{\mathrm{ULA}}$ exceeds $50\%$, the proposed closed-form Airy beam design maintains a robust and stable SE even under extreme blockage up to $90\%$. This consistent superiority, observed regardless of whether the obstacle is near the transmitter or receiver, proves that our closed-form synthesis adaptively generates the curved trajectory to ensure high-capacity performance in complex quasi-LoS environments.

\subsection{Airy Beam Design for UPA Systems}

    \begin{figure}[t]
    \centering
    \includegraphics[width=0.8\linewidth]{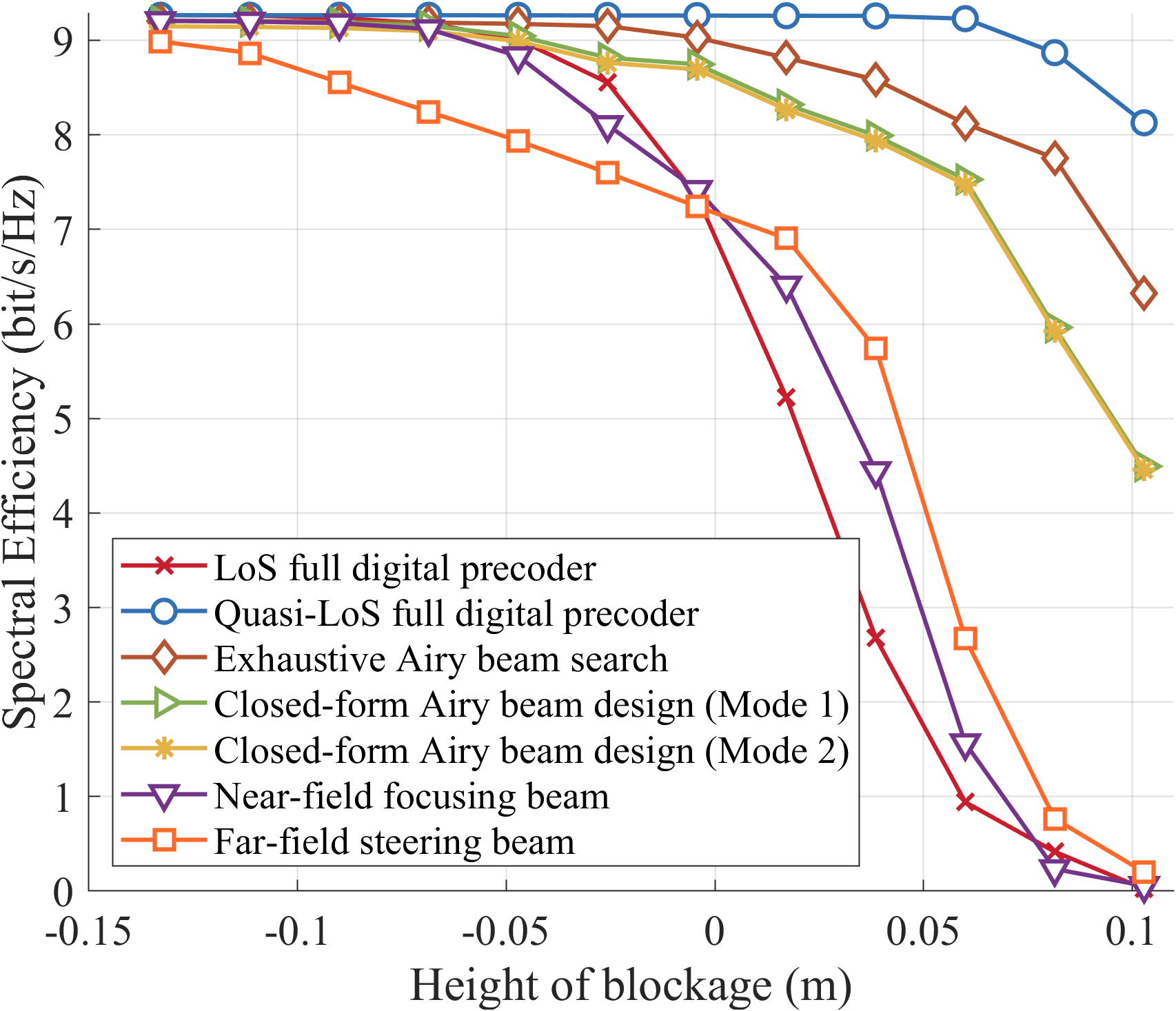} % 使用 \linewidth 自适应栏宽
    \caption{Spectral efficiency versus height of blockage in UPA system. $D=3$~m, $L=1.5$~m. $N_t=N_r=16\times16$, $d=4\lambda$.}
       \label{fig:UPA_SE}
    \end{figure}

\begin{figure}[t]
    \centering
    \includegraphics[width=0.8\linewidth]{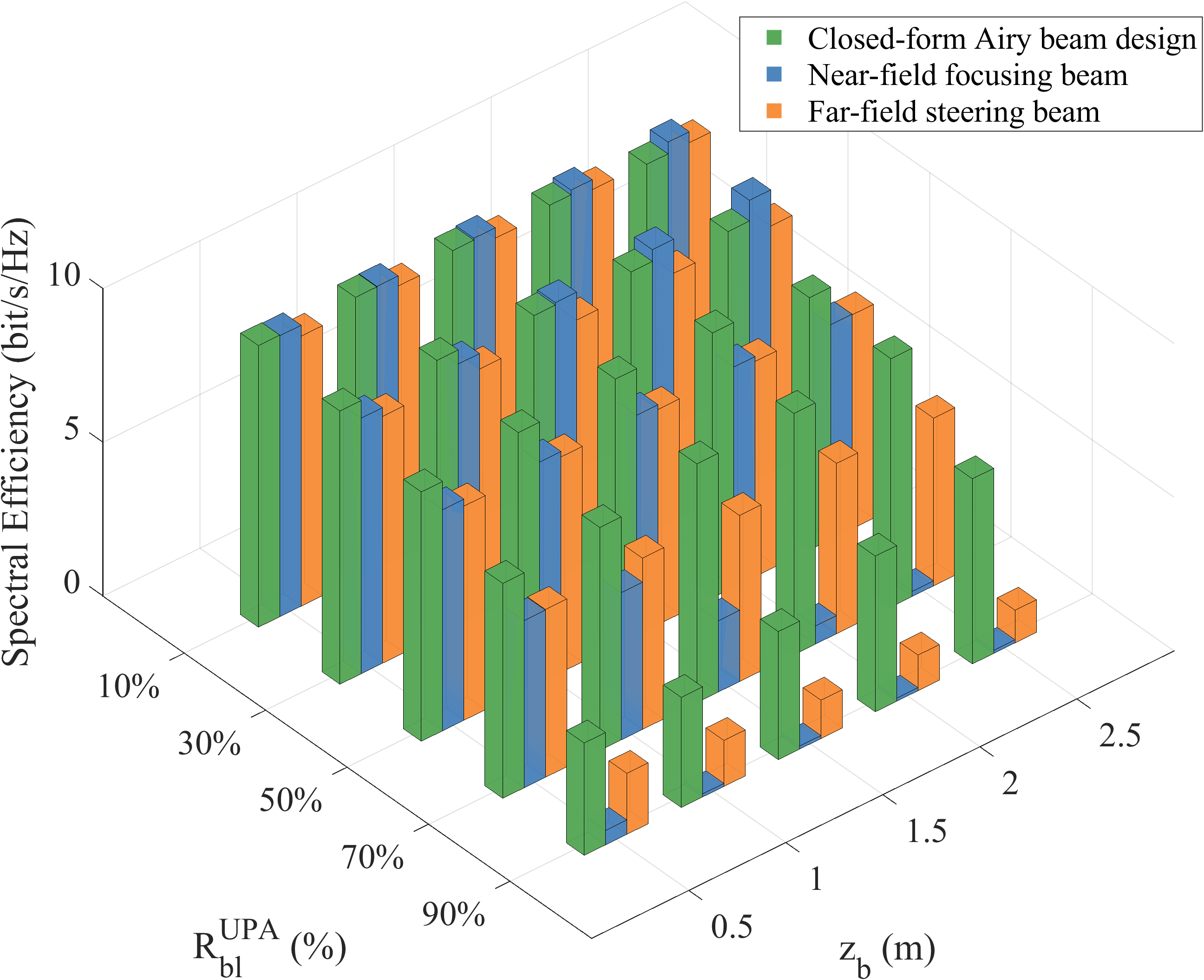} % 使用 \linewidth 自适应栏宽
    \caption{Spectral efficiency versus different position of blockage for different beamforming schemes in UPA system.}
       \label{fig:UPA_2D}
    \end{figure}

We further evaluate the performance of Airy beam in UPA systems. To demonstrate the 3D propagation characteristics, the electric field distributions for UPA beamforming under a $65.8\%$ blockage ratio are presented in Fig. \ref{fig:UPA_beam_compare}. As shown in Fig. \ref{fig:UPA_Non-blocked precoder}, Fig. \ref{fig:UPA_Focusing beam} and Fig. \ref{fig:UPA_Steering beam}, the main lobes of conventional schemes are almost entirely blocked by the obstacle, resulting in severe energy leakage. Conversely, the Airy beams in Fig. \ref{fig:UPA_Exhaustive Airy beam search}, Fig.~\ref{fig:UPA_Closed-form Airy beam design (mode 1)} and Fig. \ref{fig:UPA_Closed-form Airy beam design (mode 2)} successfully synthesize curved trajectories in 3D space. It is evident that the Airy main lobe can bend over the obstacle edge to reach the receiver. The small gap between the closed-form patterns and the exhaustive search result confirms the accuracy of our closed-form phase synthesis without the need for high-complexity iterative calculation.

As illustrated in Fig.~\ref{fig:UPA_SE}, we evaluate the spectral efficiency for UPA systems under varying blockage heights with the same benchmarks. The blockage is moved along the x-axis and $d_{py}$ remain 0.1~m. Similar to the ULA case, conventional Gaussian-based schemes experience a sharp performance decline once the blockage ratio exceeds $50\%$. In contrast, both proposed operation modes of the closed-form Airy beam design exhibit superior resilience. Specifically, the hybrid focusing-Airy mode (Mode~1) and dual Airy mode (Mode~2) achieve near identical performance, owning the negligible gap with the exhaustive search upper bound. This validates that our decoupled design strategy effectively leverages the 2D aperture to circumvent obstructions. In the high-blockage regime ($R_{bl}^{\mathrm{UPA}} > 50\%$), the proposed closed-form Airy beam design demonstrates a significant superiority, achieving an average spectral efficiency gain of approximately 4.35~bit/s/Hz, 3.77~bit/s/Hz and 3.20~bit/s/Hz, compared to the LoS full digital precoder, near-field focusing beam and far-field steering beam, respectively.

% the proposed closed-form designs provide significant SE gains over the non-blocked SVD precoder (quasi-LoS), focusing beam, and steering beam, respectively.

Finally, Fig. \ref{fig:UPA_2D} presents a comprehensive 3D comparison of SE across various obstacle distances $z_b$ and blockage ratios $R_{bl}^{\mathrm{UPA}}$. The results demonstrate the strong generalizability of the proposed UPA design. Regardless of the position of the blockage, the Airy beam maintains a robust and high SE level even under extreme $90\%$ obstruction, where conventional focusing and steering beams are ineffective. This consistent superiority across the entire LoS region proves that the proposed closed-form Airy beamforming is a reliable and computationally efficient solution for maintaining high-capacity THz links in complex 3D dynamic environments.

% \begin{figure*}[t]
%     \centering
%     \subfigure[$N_t=256,N_r=64$.]{\includegraphics[width=0.45\linewidth]{closed-form_Airy/images/simulation/UPA/UPA_SE_height_MISO.png} \label{fig:UPA_x}}
%      \subfigure[$N_t=256,N_r=256$.]{\includegraphics[width=0.45\linewidth]{closed-form_Airy/images/simulation/UPA/UPA_SE_height.png}  \label{fig:UPA_y}}
%    \caption{Validation of the closed-form trajectory equation across various curvature parameters $B$. ($N_t=256$, $D_{arr}=0.2732$ m, $d=\lambda/2$ at $140\text{ GHz}$, $F=0.5\text{ m}$).}
%    \label{fig:UPA_design}
% \end{figure*}

\section{Conclusion}
\label{sec:conclusion}
In this paper, we have investigated the quasi-LoS beamforming problem with Airy beam in THz UM-MIMO systems. To address this issue, we give a closed-form phase profile solution for Airy beam design, enabling the direct calculation of the curving, distance and angle parameters based solely on the position of the receiver and the blockage. Simulation results validate the generalizability and robustness of the proposed framework. Specifically, in ULA systems, when the blockage ratio exceeds 50\%, the proposed closed-form Airy beam design demonstrates significant superiority, by achieving spectral efficiency gains of approximately 4.63~bit/s/Hz, 4.03~bit/s/Hz and 2.33~bit/s/Hz over the traditional LoS full digital precoder, near-field focusing beam and far-field steering beam, respectively. Furthermore, in UPA systems, the two proposed modes exhibit similar performance and achieve SE gains of approximately 4.35~bit/s/Hz, 3.77~bit/s/Hz and 3.20~bit/s/Hz compared to the traditional LoS full digital precoder, near-field focusing beam and far-field steering beam. Both in ULA and UPA systems, the proposed closed-form methods show a negligible gap compared to that of the exhaustive beam search.

In summary, by enabling the direct calculation of optimal wavefront parameters with negligible computational complexity, this closed-form Airy beam design method obviates the latency of iterative beam searching while avoiding the extensive training overhead and limited interpretability inherent in learning-based methods. Therefore, it serves as a reliable and efficient solution for maintaining high-rate THz links in generic quasi-LoS scenarios.

\appendices

% \section 会自动生成 "Appendix A: Title" 的格式
% IEEE模板会自动将其居中并使用 Small Caps (小型大写字母)
\section{Derivation of Equation \eqref{eq:need_appendix}} \label{appendix:A}
To facilitate the integration in \eqref{eq:need_appendix}, we let $x_0=u+\delta$ and aim to eliminate the quadratic term in the expansion of $\frac{A}{3}(u+\delta)^3+C_2(u+\delta)^2+C_1(u+\delta)$. By expanding this expression, we identify the cubic term as $\frac{A}{3}u^3$, the quadratic term as $(A\delta+C_2)u$, the linear term as $(A\delta^2+2C_2\delta+C_1)u$, and the constant term as $\frac{A}{3}\delta^3+C_2\delta^2+C_1\delta$.

Setting the quadratic coefficient to zero yields $\delta=-\frac{C_2}{A}$. Consequently, the exponential term simplifies to
\begin{equation}
    \frac{A}{3}u^3+\left(C_1-\frac{C_2^2}{A}\right)u+\frac{2C_2^3}{3A^2}-\frac{C_1C_2}{A}.
\end{equation}
Next, we introduce $u=\gamma\tau$ with $\gamma=A^{-1/3}$ such that $\frac{A}{3}u^3=\frac{1}{3}\tau^3$. Since $du=\gamma d\tau$, the integral $I$ is reformulated as
\begin{align}
    I&=\gamma\int_{-\infty}^{+\infty}\exp\bigg[j\left(\frac{1}{3}\tau^3\right)+\left(C_1-\frac{C_2^2}{A}\right)\gamma\tau \nonumber\\&+\frac{2C_2^3}{3A^2}-\frac{C_1C_2}{A}\bigg]d\tau \nonumber \\
    &=\gamma e^{j(\frac{2C_2^3}{3A^2}-\frac{C_1C_2}{A})}\int_{-\infty}^{+\infty}e^{j(\frac{1}{3}\tau^3+(C_1-\frac{C_2^2}{A})\gamma\tau )}d\tau \nonumber \\
    &=\gamma e^{j(\frac{2C_2^3}{3A^2}-\frac{C_1C_2}{A})}2\pi\mathbf{Ai}\left[\left(C_1-\frac{C_2^2}{A}\right)\gamma\right].
\end{align}
The argument $\xi$ of the Airy function $\mathbf{Ai}(\cdot)$ is derived as
\begin{align}
    \xi
    %&=\left[-\frac{2\pi}{\lambda}\sin{\theta}-\frac{2\pi}{\lambda z}x-\frac{\frac{\pi^2}{\lambda^2}(\frac{1}{z}-\frac{1}{\widetilde{F}})^2}{(2\pi B)^3}\right]\frac{1}{2\pi B} \nonumber \\
    &=-\frac{\sin{\theta}}{\lambda B}-\frac{1}{\lambda z B}x-\frac{(\frac{1}{z}-\frac{1}{\widetilde{F}})^2}{16\lambda^2\pi^2B^4}.
\end{align}
Thus, the integrand $I$ and the final electric field profile $E(x,z)$ are expressed respectively as
\begin{equation}
    I=\frac{1}{B}\exp\left\{\frac{2(\frac{\pi}{\lambda}(\frac{1}{z}-\frac{1}{\widetilde{F}}))^3}{3(2\pi B)^6}+\frac{(\sin{\theta}+\frac{2\pi}{\lambda z}x)\frac{\pi}{\lambda}(\frac{1}{z}-\frac{1}{\widetilde{F}})}{(2\pi B)^3}\right\}\mathbf{Ai}(\xi)
\end{equation}
\begin{equation}
    E(x,z)=\frac{e^{jkz}}{j\lambda z B}e^{j\frac{\pi}{\lambda z}x^2}e^{j(\frac{2C_2^3}{3A^2}-\frac{C_1C_2}{A})}\mathbf{Ai}(\xi).
\end{equation}

\newpage

\end{document}